 \documentclass{aa}  
\usepackage{graphicx}
\usepackage[varg]{txfonts}
\usepackage{natbib}
\bibpunct{(}{)}{;}{a}{}{,} 

\usepackage{color}
\usepackage{longtable}
\begin{document}

    \title{\textit{Herschel} far-infrared observations of the Carina Nebula complex\thanks{The \textit{Herschel} data described in this paper
    have been obtained in the open time project {\tt OT1\_tpreibis\_1} (PI: T.~Preibisch). \textit{Herschel}
    is an ESA space observatory with science instruments provided by European-led Principal Investigator
    consortia and with important participation from NASA.},\thanks{Tables A.1, B.1, C.1, and D.1 are only available in electronic form at the CDS.}
    }
    \subtitle{II: The embedded young stellar and protostellar population}

   \author{B.~Gaczkowski\inst{1} \and T.~Preibisch\inst{1} \and T.~Ratzka\inst{1}  \and V.~Roccatagliata\inst{1} \and H.~Ohlendorf\inst{1} \and H.~Zinnecker\inst{2,3}
          }

   \institute{Universit\"ats-Sternwarte M\"unchen, 
        Ludwig-Maximilians-Universit\"at,
          Scheinerstr.~1, 81679 M\"unchen, Germany;
	  \email{bengac@usm.uni-muenchen.de}
	\and
 Deutsches SOFIA Institut, Universit\"at Stuttgart,
 Pfaffenwaldring 31, 70569 Stuttgart, Germany
    \and
NASA-Ames Research Center,
MS 211-3, Moffett Field, CA 94035, USA
                } 

 \titlerunning{\textit{Herschel} detected protostars in the Carina Nebula complex}

   \date{Received 18 June 2012;  accepted 08 November 2012}

 
   \abstract
{The Carina Nebula represents one of the largest and most active star forming regions
known in our Galaxy. It contains numerous very massive ($M \ga 40\,M_\odot$) stars that
strongly affect the
surrounding clouds by their ionizing radiation and stellar winds.}
{Our recently obtained \textit{Herschel} PACS \& SPIRE far-infrared maps cover the full
area $(\approx8.7~{\rm deg}^2)$ of the Carina Nebula complex and
reveal the population of deeply embedded young stellar objects, most of which
are not yet visible in the mid- or near-infrared.}
{We study the properties of the 642 objects that are independently detected as point-like sources
in at least two
of the five \textit{Herschel} bands. For those objects that can be identified with
apparently single \textit{Spitzer} counterparts, we use radiative transfer models
to derive information about the basic stellar and circumstellar parameters.}
{We find that about 75\% of the \textit{Herschel}-detected YSOs are Class~0 protostars.
The luminosities of the \textit{Herschel}-detected YSOs with SED fits are restricted to values
of $\le 5400\,L_\odot$, their masses (estimated from the  radiative transfer modeling)
range from $\approx 1\,M_\odot$ to $\approx 10\,M_\odot$.
Taking the observational limits into account and extrapolating the observed number of
\textit{Herschel}-detected protostars over the stellar initial mass function suggest that
the star formation rate of the CNC is $\sim 0.017\,M_\odot/{\rm year}$.
The spatial distribution of the
\textit{Herschel} YSO candidates is highly inhomogeneous and does \textit{not} follow
the distribution of cloud mass. Rather, most \textit{Herschel} YSO candidates are found
at the irradiated edges of clouds and pillars. The far-infrared fluxes of the famous object
$\eta$~Car are about a factor of two lower than expected from observations with the
\textit{Infrared Space Observatory} obtained 15 years ago;
this difference may be a consequence of dynamical changes in the circumstellar dust in the
Homunculus Nebula around $\eta$~Car.}
{The currently ongoing star formation process forms only low-mass and intermediate-mass stars, but no
massive ($M \ga 20\,M_\odot$) stars. The characteristic spatial configuration of the YSOs provides
support to the picture that the formation of this latest stellar generation is triggered by
the advancing ionization fronts.}

 
   \maketitle
%

\section{Introduction}

Most stars in our Galaxy form in giant molecular clouds, as parts of rich stellar clusters 
or associations, containing high-mass ($M>20\,M_\odot$) stars. 
Recent investigations have shown that also our solar system formed close to massive stars, 
which had important influences on the early evolution of the solar nebula \citep[e.g.,][]{solar-system-birth}. 
The presence of hot and luminous O-type stars leads to physical conditions that are 
very different from those in regions like Taurus where only low-mass stars form. High-mass stars create H\,II regions due to their strong UV radiation, generate wind-blown bubbles, and explode as supernovae. The negative feedback from high-mass stars destroys their surrounding molecular clouds \citep[see e.g.,][]{massive-stars-cloud-dispersal,radiative-feedback-cluster} and can halt further star formation. Young stellar objects (YSOs) may also be affected directly by the destructive UV radiation from nearby massive stars \citep[][]{ysos-disk-dispersal-low-final-mass} that can disperse their disks, leading to a deficit of massive
 disks \citep[][]{ysos-disk-evaporation}. However, massive star feedback can also have positive effects and lead to triggered star formation. Advancing ionization fronts and expanding superbubbles compress nearby clouds, increasing their density and causing the collapse of deeply embedded cores. This leads to new star formation.\\

The Carina Nebula (NGC~3372; see \cite{carina-overview} for an overview) is a perfect location in which to study massive star formation and the resulting feedback effects. Its distance is well
 constrained to 2.3~kpc \citep[][]{carina-distance} and its extent is about 80~pc (corresponding to $2\degr$ on the
 sky). The Carina Nebula complex (CNC hereafter) represents the nearest southern region with a large massive stellar
 population. Among the 65 known O-type member stars \citep[][]{carina-o-stars} are some of the most massive
 ($M>100\,M_\odot$) and luminous stars in our Galaxy. These include the famous Luminous Blue Variable $\eta~{\rm Carinae}$ \citep[see][]{carina-homonucleus, carina-o-stars}, the O2 supergiant HD 93129A \citep[see][]{carina-hd93129a} with about $120\,M_\odot$, and also four Wolf-Rayet stars \citep[see][]{carina-wolf-rayet-2, carina-wolf-rayet-1}. Most of
 the very massive stars are gathered in several open clusters, including Trumpler 14, 15 and 16. For these clusters, ages between $\lesssim3~{\rm Myr}$ (Tr~14 and 16) and $\approx5-8~{\rm Myr}$ (Tr~15) have been found \citep[see][]{carina-tr14-age, carina-hawki}.
The region contains more than $10^5\,M_\odot$ of gas and dust
 \citep[see][]{carina-overview,carina-laboca,carina-herschel-clouds}. 

Recent sensitive infrared, sub-mm, and radio observations 
showed clearly that the Carina Nebula complex is a site of ongoing star formation. 
First evidence for active star formation in the Carina Nebula was found by \cite{carina-sf-old} who 
identified four young stellar objects from near-infrared observations with IRAS. 
\cite{carina-south-pillars-ysos} characterized 38 objects of the Red MSX Source (RMS) mid-infrared 
survey in the area of the CNC as massive YSO candidates. From their \textit{Spitzer} survey of an 
$\approx0.7\,{\rm deg}^2$ area in the South pillars region of the Carina Nebula, 
\cite{carina-south-pillars-spitzer} classified 909 sources as YSO candidates. Furthermore, 
40 Herbig-Haro jets have been discovered by \cite{carina-south-pillars-hst} through a deep 
Hubble Space Telescope (HST) H$\alpha$ imaging survey. The driving sources of many of these jets
were recently revealed and analyzed by \cite{carina-jets-henrike}. 
\cite{carina-south-pillars-spitzer-protostars-intermediate} published a catalog of 1439 YSO
 candidates (PCYC catalog) based on mid-infrared excess emission detected in the \textit{Spitzer} data. 
A recent deep wide-field X-ray survey revealed $10\,714$ young stars in a $\sim 1.5$~square-degree area
centered on the Carina Nebula
\citep{carina-chandra-townsley, carina-chandra}. The X-ray, near-, and mid-infrared observations 
provided comprehensive information about the (partly) \textit{revealed} young stellar population of stars 
(i.e.,~Class~I protostars and T~Tauri stars, with ages between $\sim 10^5~{\rm yr}$ and a few Myrs).
However, no systematic investigation of the youngest, deeply embedded population of the currently 
forming protostars was possible so far, because no far-infrared data with sufficient sensitivity 
and angular resolution to detect a significant fraction of the embedded protostars existed until now.

 The \textit{Herschel} far-infrared observatory \citep[][]{herschel} is currently observing
 many star forming regions \citep[see e.g.,][]{herschel-hobys,herschel-gould-belt,herschel-bubble-hII-regions}
and is very well suited to detect deeply embedded protostars 
\citep[][]{herschel-vela-c,herschel-aquila-rift,herschel-lmc-protostar}, which cannot (yet) be seen 
in the mid- and near-infrared. We used \textit{Herschel} to map the entire Carina Nebula complex 
($\approx10.2~{\rm deg}^2$)
with PACS and SPIRE. A general description of these \textit{Herschel} observations and first results 
about the global properties of the clouds have been presented in \cite{carina-herschel-clouds}.

\begin{figure*}[!htb]
\centering
\includegraphics[width=\linewidth, height=\textheight, keepaspectratio]{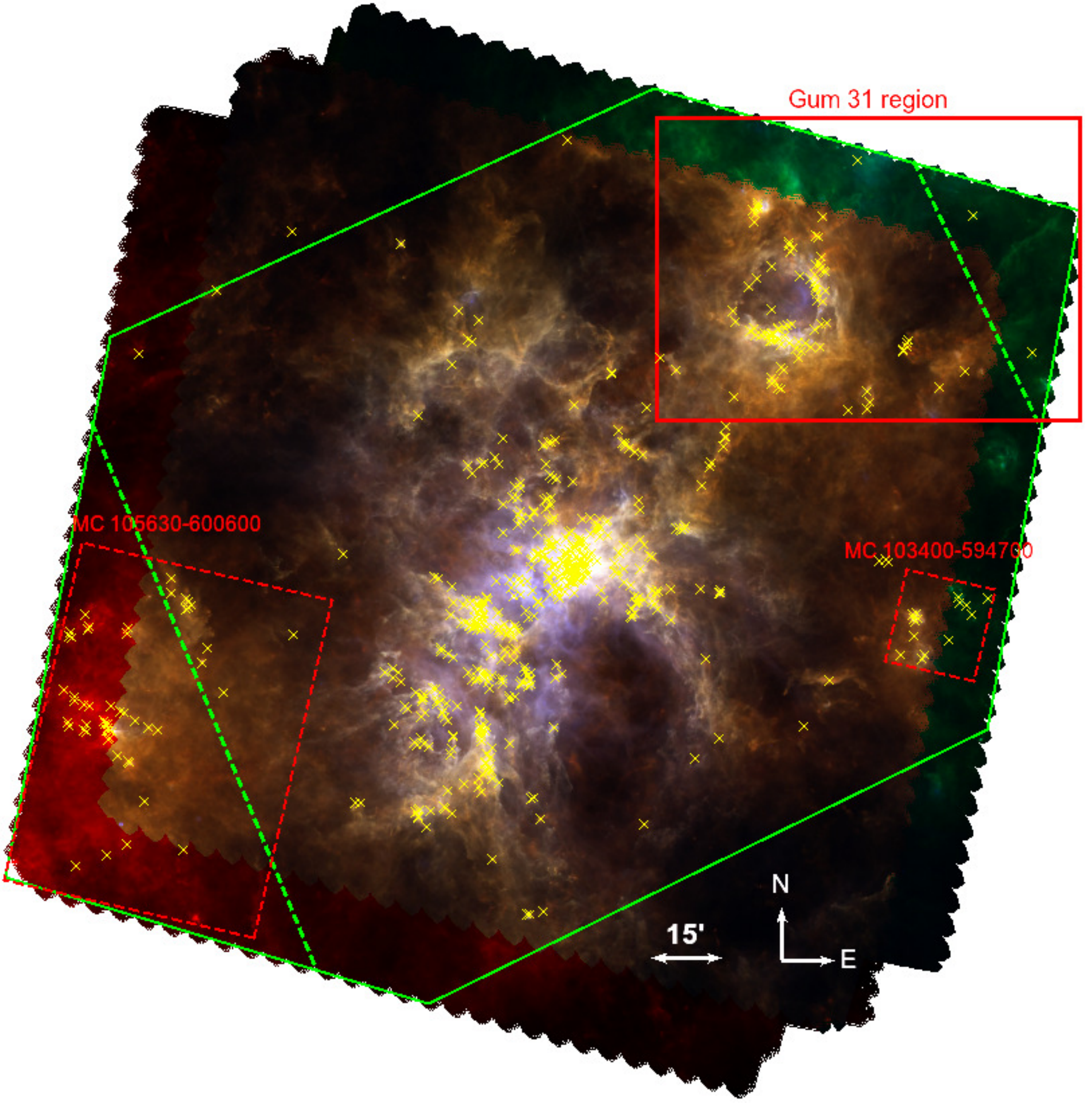}
\caption{\textit{Herschel} three color composite of the CNC maps at $70\,\mu$m (blue), $160\,\mu$m (green) and $250\,\mu$m (red). The yellow 'x' mark the positions of our  642 \textit{Herschel} point-like sources detected in at least two bands. The solid red box delineates the Gum~31 region that will be discussed in a separate paper. The two dashed red boxes delineate two clouds that are not part of the Carina Nebula and will also be discussed in a separate paper. The green polygon marks the region of the Carina Nebula that is also covered by the \textit{Spitzer} IRAC maps. The two green dashed lines within the polygon mark the borders of the region for which our \textit{Spitzer} photometry was obtained. Because the field-of-views of PACS and SPIRE are shifted to each other, there are regions that are only covered by one instrument, i.e. the red region in the south-west is only part of our SPIRE maps, and the green region in the north-east is only part of our PACS maps, respectively.}
\label{img:analyzed-region-cut}
\end{figure*}

This paper focuses on the detection and investigation of the point-like sources in the \textit{Herschel} 
maps. We present a catalog of  642 far-infrared point-like sources detected (independently) in at least
two of the five  \textit{Herschel} bands. 
For the 483 objects located in the central $\approx8.7~{\rm deg}^2$ region of the Carina Nebula
(see Fig.~\ref{img:analyzed-region-cut}), we search for counterparts
in the \textit{Spitzer} mid-infrared maps and construct their spectral energy distributions (SEDs) 
from $\sim 1\,\mu$m to $500\,\mu$m. Modeling of these SEDs provides information about basic
stellar and circumstellar parameters of the YSOs.
The properties of the  92 \textit{Herschel} point-like sources in the region around the 
Gum~31 nebula, at the north-western part of our maps, will be presented in a separate paper 
(Ohlendorf et al. 2012, A\&A submitted).

\section{Observations and data reduction}

We present \textit{Herschel} far-infrared data of the CNC and complement it with \textit{Spitzer}, \textit{2MASS}, and \textit{WISE} data to obtain information in the shorter wavelengths. This allows the construction of SEDs over a wider wavelength regime than \textit{Herschel} alone.

\subsection{\textit{Herschel} far-infrared maps}

The Carina Nebula complex was observed by the \textit{Herschel} satellite on December 26th, 2010. The maps obtained cover an area of $3.2\degr\times3.2\degr$ corresponding to a physical region of 128~pc $\times$ 128~pc at the distance of the CNC, i.e. including the full extent of the
 complex. The CNC was simultaneously imaged in five different wavelengths,
 using the two on-board photometer cameras PACS \citep[][]{pacs} at 70 and $160\,\mu$m and SPIRE \citep[][]{spire} at 250, 350, and $500\,\mu$m. Two orthogonal scan maps were obtained by mapping in the parallel fast scan mode with a velocity of $60\arcsec/{\rm s}$. The total observing time was 6.9 hours. 
\smallskip

The data reduction was performed with the HIPE v7.0
\citep[][]{hipe} and {\tt SCANAMORPHOS} v10.0 \citep[][]{scanamorphos} software packages. From level 0.5 to 1 the PACS data were reduced using
the {\tt L1\_scanMapMadMap} script in the photometry pipeline in
HIPE with the version 26 calibration tree. The level~2 maps
were produced with {\tt SCANAMORPHOS} with standard options
for parallel mode observations, including turnaround data. The pixel-sizes for the two PACS maps at 70 and $160\,\mu$m were chosen as $3.2\arcsec$ and
$4.5\arcsec$, respectively, as suggested by \cite{higal}.

The level 0 SPIRE data were reduced with an adapted version
of the HIPE script {\tt rosette\_obsid1\&2\_script\_level1} included
in the {\tt SCANAMORPHOS} package. Here the version 7 calibration tree was used. The final maps were produced
by {\tt SCANAMORPHOS} with standard options for parallel
mode observations (turnaround data included). The pixel-sizes for the three SPIRE maps at 250, 350 and $500\,\mu$m were chosen as $6\arcsec$, $8\arcsec$ and $11.5\arcsec$, respectively. The angular resolutions of the maps are $5\arcsec$, $12\arcsec$, $18\arcsec$, $25\arcsec$, and $36\arcsec$ for the 70, 160, 250, 350, and $500\,\mu$m band, respectively. At the distance of the CNC this corresponds to physical scales from 0.06 to 0.4~pc.

\smallskip

\subsection{Spitzer IRAC data}

We retrieved the available data from the \textit{Spitzer} Heritage Archive
(PI: Steven R. Majewski; Program-ID: 40791) and assembled the basic calibrated data into wide-field ( $\approx 2.6\degr \times 3.0\degr$)
mosaics that cover nearly the full extent of the Carina Nebula.
As illustrated in  Fig.~\ref{img:analyzed-region-cut}, the \textit{Spitzer} IRAC maps
cover $\approx87\%$ of the area of our \textit{Herschel} maps.
We performed point source detection and photometry in these
IRAC mosaics, using
the Astronomical Point source EXtractor (APEX) module of the MOsaicker and Point source
EXtractor package \citep[MOPEX; ][]{spitzer-mopex}, as described in \cite{carina-jets-henrike}.

The entire \textit{Spitzer} point-source catalog contains 569\,774 objects;
548\,053 of these are located in the area covered by our \textit{Herschel} maps.
The fluxes of the faintest objects in our point-source catalog range from
$\approx 0.1$~mJy in the IRAC~1 and IRAC~2 maps, over $\approx 0.2$~mJy in IRAC~4,
to  $\approx 0.4$~mJy in IRAC~3. However, the detection limit is a strong function
of location in the maps, because large parts of the mosaics are pervaded by
very strong and highly inhomogeneous diffuse emission that reduces the
local source detection sensitivity considerably.

In order to estimate typical values for the completeness limits of our Spitzer IRAC catalog
across the field,  we inspected the flux-distributions and determined the
points  where the flux histograms start to deviate from a power law-shape.
This was found to occur at $\approx 1.1$~mJy for IRAC~1, $\approx 0.4$~mJy for IRAC~2,
$\approx 1.2$~mJy for IRAC~3, and $\approx 1.2$~mJy for IRAC~4.
These values can be regarded as typical \textit{average} completeness limits;
at locations of particularly bright [faint]
nebulous emission the sensitivity can be considerably poorer [better].
\medskip

We also retrieved a mosaic map of the CNC from the available \textit{Spitzer} MIPS archive (PI: Jeff Hester; Program-ID: 20726). However, since a large fraction of the MIPS map
is saturated by the very strong diffuse emission in the central
parts of the Carina Nebula, no attempt
was made to construct a photometric source catalog.

\section{Construction of the catalog of point-like Herschel sources}

\subsection{Source detection and photometry}

\begin{figure*}[!htb]
\centering
\includegraphics[width=\linewidth, height=\textheight, keepaspectratio]{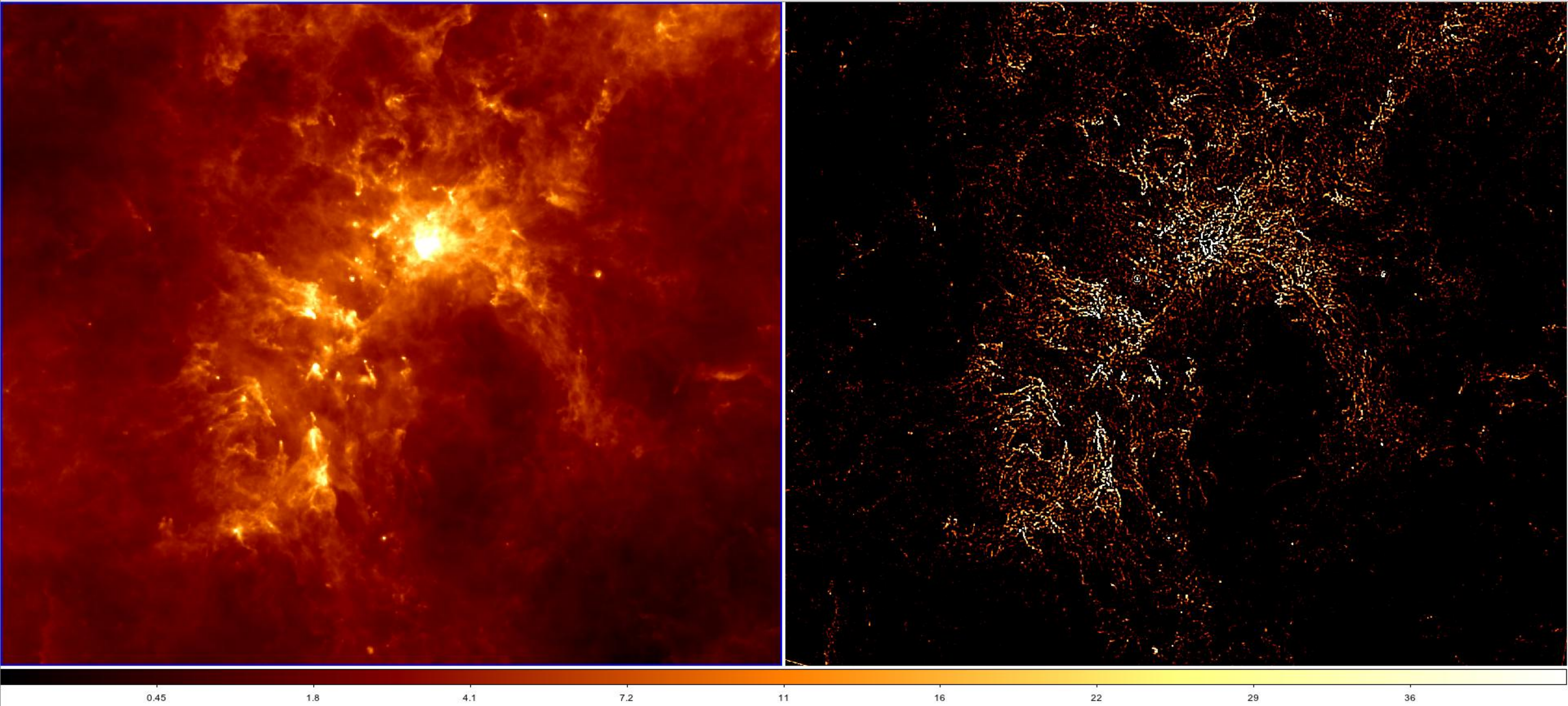}
\caption{Second derivative image (on the right) of the SPIRE $250\,\mu$m map (on the left). Diffuse emission is dampened out and point-like sources are easily detectable by eye.}
\label{img:cutex-deriv}
\end{figure*}

The point source detection and photometry in the five \textit{Herschel} maps was carried out with {\tt CuTEx} \citep[][]{cutex}, a software package developed especially for maps with complex background. It was developed and extensively tested by the \textit{Herschel} infrared Galactic Plane Survey ({Hi-GAL}) team and is used for all point-source detection and photometry in the Hi-GAL project \citep[][]{higal-maps}. It calculates the second order derivatives of the signal map in four directions (x, y and their
 diagonals). Point-like sources produce steep brightness gradients compared to their surrounding background and can therefore be identified in the derivative of the map,
 i.e. curvature of the brightness distribution. Point sources should also have a
 similarly steep brightness gradient in all directions; this helps to distinguish them from elongated structures like filaments, which are a very prominent feature in the \textit{Herschel}
 images. An image with an average of all four derivatives is shown for our SPIRE
 $250\,\mu$m map in Fig.~\ref{img:cutex-deriv}. The differentiation dampens the background
 emission, enhancing the visibility of the point-like sources with respect to the original map. This allows to apply thresholding methods to detect the source peaks. A detection is
 considered significant when a certain curvature threshold is exceeded in all differentiation
 directions. For the photometry the routine assumes that the source brightness distribution can be approximated by a 2-dimensional elliptical Gaussian profile with variable size (FWHM) and orientation (position angle). To the varying background, the
  routine simultaneously fits an additional planar plateau at variable inclination
  and direction. This is done by cutting a limited fitting window around the source and estimating the background within this window. The fitted profiles are integrated to obtain
  the fluxes of the point-like sources.
\smallskip

The best values for the detection parameters were chosen by visual inspection of
 the images. Our aim was that, on one hand, the faintest sources seen by eye
 were detected by the algorithm. On the other hand, care has been taken that nebulous extended structures 
and artifacts are excluded by the algorithm. We used a conservative approach that minimizes the number of spurious detections and excludes even slightly extended features such as nebular knots as far as possible. The parameters for the source detection step and the final number of point-like sources detected in each of the five bands are
listed in Table~\ref{tab:cutex-detect-results}.

For the photometry we set the {\tt /backgfit} keyword 
in all maps to obtain a second order background fit. The other parameters were left at their default values. 
For some sources in the PACS $70\,\mu$m map, {\tt CuTEx} assumed too small values for the PSF size.
These cases were identified by visual inspection and this problem was solved by enlarging their PSF size values.

\begin{table*}[!htb]
\centering
\caption[]{Detection parameters and final source counts of {\tt CuTEx} in our \textit{Herschel} maps. The second, third and fourth column gives the final detection parameters used in {\tt CuTEx}. The fifth column shows the number of detected point-like sources and the last one shows the number of sources in each band that are detected only in this band.}
\label{tab:cutex-detect-results}
\begin{tabular}{r c c c c c c c c}
\hline\hline
\noalign{\smallskip}
 Band & pixel size & FWHM & beam size & {\tt PSFPIX} & {\tt NPIX} & {\tt thr} & \# Detections & \# no detections in other bands\\ $[\mu$m] & [arcsec/pixel] & [arcsec] &  [$\rm{arcsec}^2$]  \\
\noalign{\smallskip}
\hline
\noalign{\smallskip}
   70 & 3.2  & 5  & 28   & 1.56  & 4.0 & 9.5 & 454 & 168 \\
  160 & 4.5  & 12 & 148  & 2.70  & 4.0 & 8.5 & 552 & 72 \\
  250 & 6.0  & 18 & 361  & 2.98  & 4.0 & 6.5 & 650 & 104 \\
  350 & 8.0  & 25 & 708  & 3.13  & 4.0 & 6.0 & 471 & 36 \\
  500 & 11.5 & 36 & 1445 & 3.11  & 4.0 & 5.5 & 253 & 21 \\
\noalign{\smallskip}
\hline
\end{tabular}
\end{table*}

\subsection{The Herschel point source catalog}\label{sec:The Herschel point source catalog}

Our final \textit{Herschel} catalog was constructed in a very conservative way
to provide a reliable and objective sample of point-like sources.
As any source catalog based on maps with strong and
highly spatially inhomogeneous background emission, our \textit{Herschel} source lists
for the individual maps may contain some number of spurious detections.
In order to exclude spurious detections as far as possible,
we included in our final catalog only those sources are detected independently in
at least two different \textit{Herschel} maps.\\
For this, the lists of point-like sources detected in each individual \textit{Herschel} band
were matched. The matching radius was chosen as $15\arcsec$ to account for the different
angular resolution in the five bands. Starting with the shortest wavelength of
$70\,\mu$m, i.e.~the band with the best angular
 resolution, each detected source at this wavelength (parent source in the following) is
 checked for a match in every other \textit{Herschel} band.  A match is found if the source position
 lies within the matching radius  of the parent source. If more
 than one match is found for the parent source within the matching radius, the closest is
 chosen. If, on the other hand, several parent sources have the same match in another band,
 this match is assigned to the closest parent source. This procedure is then
 repeated subsequently for the longer wavelengths. This procedure preserves for every source the
 coordinates derived from the map with the shortest wavelength in which it was detected,
i.e.~best angular resolution.\\
In the full area of our \textit{Herschel} maps,  642 point-like sources were independently
detected in at least two \textit{Herschel} bands. The final statistics of this matching process
can be found in Table~\ref{tbl:sources-matching-results}.\\
The photometric catalog of these sources can be found in the appendix (Table~\ref{tbl:phot-fluxes-631}).\\ \\
While the use of such restrictive criteria leads to a very reliable catalog, it also
automatically implies that a considerable number of true sources will be rejected,
e.g.~because small distortions of the PSF prevent an classification as ``point-like'' by
the detection software.
Our requirement of detection in at least two bands will also
automatically remove many faint sources for which only the flux in the band
closest to the peak of their spectral energy distribution is above the
local detection limit. Therefore, single-band detections are not necessarily
spurious sources.\\ \\
We therefore produced a second catalog that lists the properties of
these 500 additional \textit{Herschel} source-candidates. It includes the objects
detected by {\tt CuTEx} in only one band, as well as a number of additional point-like sources
that were found by visual inspection of the maps. It can be found in the appendix (Table~\ref{tbl:single-herschel-sources}).\\ \\
Our analysis in the rest of this paper remains, however, restricted to the
sample of 642 point-like sources that were reliably detected in at least two bands.

\begin{table}[htb]
\centering
\caption[Final matching results of the point-like sources.]{Final matching results all  642 point-like sources in the five \textit{Herschel} bands.}
\label{tbl:sources-matching-results}
\begin{tabular}{r r}
\hline\hline
\noalign{\smallskip}
 \# bands &  \# Sources \\
\noalign{\smallskip}
\hline
\noalign{\smallskip}
 $\ge 2$  &  642 \\
 $\ge 3$  &  418 \\
 $\ge 4$  &  209 \\
   $= 5$  &   75 \\
\noalign{\smallskip}
\hline
\end{tabular}
\end{table}

\subsection{Sensitivity limits of the Herschel point source catalog}\label{sec:Herschel sensitivity}

According to the \textit{Herschel} documentation\footnote{SPIRE PACS Parallel Mode Observers' Manual (HERSCHEL-HSC-DOC-0883, Version 2.1, 4-October-2011)}, the theoretical $3\sigma$ sensitivity limits of 
PACS and SPIRE parallel mode observations in fast scan mode is about  40, 100, 25, 20, and 30~mJy in the 70, 160, 250, 350, and $500\,\mu$m band, respectively. 
These values are however, only theoretical limit for isolated point sources in regions without 
significant diffuse background. 
The true sensitivity limits in fields with strong and inhomogeneous background emission, 
as present in our maps of the Carina Nebula, are considerably higher. 
Due to the very strong spatial inhomogeneity of the cloud emission in our maps,
the sensitivity cannot be precisely quantified by a single value.\\
Instead, we characterize it by two typical values, the \textit{optimum detection limit}
 and the \textit{mode of the flux distribution}.\\
The fluxes of the faintest detected sources are in the range $\approx 1 - 2$~Jy
in our maps; this provides an estimate of the detection limit in regions without
strong background emission.\\
The complex and bright background does not allow to determine a well-defined
``completeness limit'' for our maps. The modal values of the flux distributions
are at $F_{\rm 70} \sim 10~{\rm Jy}$, $F_{\rm 160} \sim 15~{\rm Jy}$,
$F_{\rm 250} \sim 10~{\rm Jy}$, $F_{\rm 350} \sim 10~{\rm Jy}$, and $F_{\rm 500} \sim 6~{\rm Jy}$.
These values can serve as a rough proxy for the
typical completeness limit across the field
(i.e.~the limit above which we expect most sources in the
survey area to be detected as point-sources).

 \begin{figure*}[!htb]
 \centering
 \includegraphics[width=\linewidth, keepaspectratio]{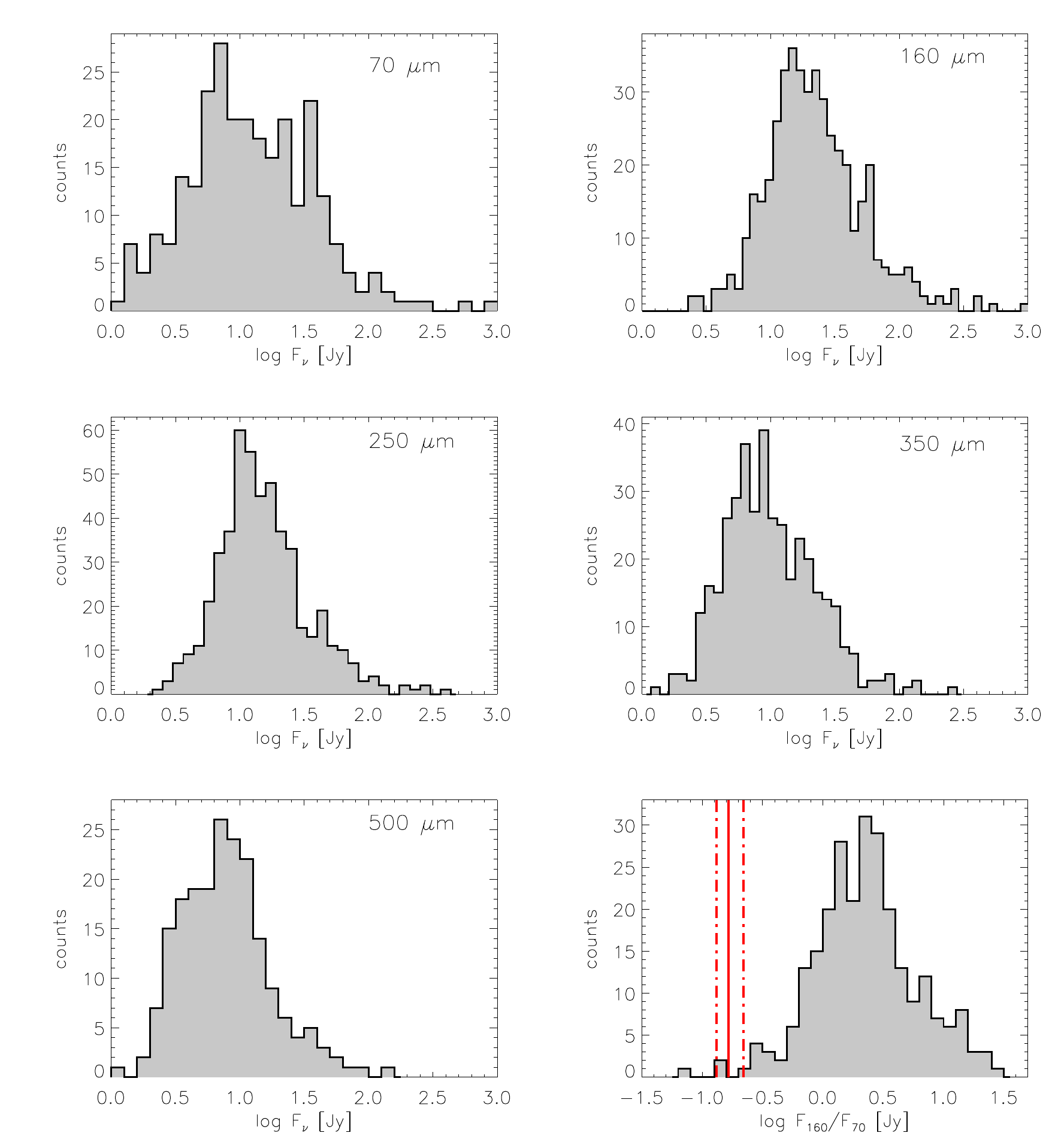} 
 \caption{Flux histograms of the \textit{Herschel} point-like sources in the CNC and Gum~31 for all five bands. In the bottom right corner a histogram of the flux ratio $F_{160}/F_{70}$ is shown. The solid red line marks the typical flux ratio for evolved stars which is $\sim6$ as determined by \cite{agb-mass-loss-herschel}. The two dot-dashed lines mark the $1\sigma$ interval of this ratio.}
   \label{img:herschel-flux-histos}
 \end{figure*}

\section{The nature of the \textit{Herschel} point-like sources}

\subsection{Young stellar objects}

The canonical model for the evolution of YSOs starts with the collapse of a pre-stellar core
and proceeds through the embedded protostellar phase, where most of the mass is
still in the circumstellar envelope, via
a pre-main sequence star with circumstellar accretion disk to a ZAMS
without (significant amounts of) circumstellar material.
Observationally, these different phases can be traced by characteristic
differences in the spectral energy distribution: according to the
often used infrared classification scheme based on the slope of the near- to mid-infrared
spectral energy distribution \citep[][]{embedded-ps-rho-oph,from-pre-to-protostar-andre1993},
Class~0 objects should represent early phases of protostars
in collapsing cloud cores;
Class I objects are more evolved protostars, but still embedded in a
relatively massive, in-falling envelope;  in
Class II objects, the young star is nearly fully assembled, but still
accreting from a circumstellar disk,
while class III objects are pre-main sequence stars that have already
dispersed their disks \citep[see][]{embedded-ps-properties}.

Although we consider only ``point-like'' \textit{Herschel} sources in this study,
it is important to keep in mind that the relatively large PSF corresponds
to quite large physical scales at the 2.3~kpc distance of the Carina Nebula.
In the PACS $70\,\mu$m map, all objects with an angular [spatial] extent of up to
$\approx 5''$ [11\,500~AU = 0.056~pc] are compact enough to appear ``point-like''.
For the SPIRE $250\,\mu$m map, the corresponding numbers are
$\approx 18''$ [41\,400~AU = 0.20~pc].
This shows immediately that (pre-stellar) cloud cores, which
have typical radii of $\la 0.1$~pc, cannot be (well) resolved in the
 \textit{Herschel} maps and appear as compact ``point-like'' sources.
This implies that YSOs in all the above mentioned stages can, in principle,
appear as point-like sources in our \textit{Herschel} maps.
However, the possibility to detect an object in a specific stage depends
strongly on its properties; as described below, many pre-stellar cores and
embedded protostars will be easily detectable, while most of the more evolved
pre-main sequence stars with disks should remain undetected.\\ \\

Another possible problem could be externally illuminated nebular knots in the clouds.
Such knots may be gravitationally unbound and thus never collapse to form stars,
but if they are compact enough, they may nevertheless appear as point-like sources
in our \textit{Herschel} maps. Such knots constitute a general problem that concerns all \textit{Herschel} maps, and is not specific to our study. From the \textit{Herschel} data alone, it is not possible to reliably distinguish gravitationally unbound knots from bound cloud cores.
However, the detection limits of our \textit{Herschel} maps imply
that all detected compact clouds must have substantial mass; at least one
solar mass (see below). Minor inhomogeneities at the surface of large-scale
clouds are thus unlikely to appear as detectable \textit{Herschel} sources.

\subsubsection{Herschel sensitivity to circumstellar matter}\label{sec:Herschel sensitivity to circumstellar matter}

The far-infrared emission from a YSO is dominated by the thermal emission
from circumstellar dust. The level of the far-infrared flux
depends on several factors; the most important ones are
(i) the amount of circumstellar material, (ii) the
spatial distribution of this material, and (iii)
the luminosity of the central YSO.
Considering our \textit{Herschel} detection limits of $\approx 1$~Jy,
we used radiative transfer models to
estimate the minimum values of the circumstellar mass and YSO luminosities
that are required for a detection in at least two of the five \textit{Herschel} bands.

\paragraph{Pre-stellar cores:}
Radiative transfer simulations with the
code and the dust model described in \cite{ice-grains-ps-spectra},
show that spherical pre-stellar cloud cores (i.e.~no internal
source of luminosity) with radii of 0.1~pc and temperatures of
$T=20$~K can be detected for cloud masses of $\ga 2\,M_\odot$.
Depending on the level of surface irradiation  by nearby hot stars,
this mass limit can decrease to $\ga 1\,M_\odot$.

Since the SEDs of pre-stellar cores drop steeply for wavelengths
shorter than $\sim 100\,\mu$m, no emission is expected to be detectable
in the \textit{Spitzer} IRAC maps.

\cite{herschel-epos-pre-high-mass-sf} presented radiative transfer models of starless cores
and protostellar cores and investigated the detectability of these two different classes of
objects. Their models showed that the SEDs of starless (i.e.~pre-stellar) cores
typically peak around $\approx200-300\,\mu$m and drop very steeply towards shorter
wavelengths. Their model fluxes at $70\,\mu$m (scaled to the distance of the CNC) are several
orders of magnitudes below our detection limits.\\
Protostellar cores, on the other hand, have much stronger fluxes at PACS wavelengths.
Guided by these results, we can thus use a detection at $70\,\mu$m as an indication
for the protostellar nature of the source, whereas \textit{Herschel} sources without
$70\,\mu$m detection could be pre-stellar cores.\\
According to this, about 50\% of our \textit{Herschel} point-like sources  are most likely
protostellar objects.

\paragraph{Embedded protostars:} 
We performed radiative transfer simulations with the
code and the dust model for protostellar envelopes
described in \cite{ice-grains-ps-spectra}.
We assumed the central protostar to be surrounded by a spherical dust envelope 
with a radius of 5000~AU and density power law
$\rho(r) \propto r^{-1.5}$. In Table~\ref{tbl:min-circu-mass} we list for 
protostars of different mass (and corresponding luminosity)
the minimum circumstellar envelope mass required for a detection in
at least two of the five \textit{Herschel} bands. 
As the luminosities of the protostars
are highly time-dependent, the values listed were chosen
considering the models of \cite{accretion-physics-onto-ys}, \cite{sf-in-orion}, 
\cite{evolution-massive-ps-via-disk-accretion},  and \cite{radiative-feedback-cluster} and  should be regarded as
``typical'' values.

\begin{table}[htb]
\centering
\caption{Minimum circumstellar envelope mass required for a detection in
at least two of the five \textit{Herschel} bands for protostars of different mass.}
\label{tbl:min-circu-mass}
\begin{tabular}{rrl}
\hline\hline
\noalign{\smallskip}
$M_{\rm protostar}$ & $L_{\rm protostar}$ & $\;\;\;\;M_{\rm env}$ \\
\noalign{\smallskip}
\hline
\noalign{\smallskip}
$1\,M_\odot$   & $30\,L_\odot$ & $\ga 0.5\,M_\odot$ \\
$2\,M_\odot$   & $90\,L_\odot$ & $\ga 0.25\,M_\odot$\\
$4\,M_\odot$   & $200\,L_\odot$ & $\ga 0.13\,M_\odot$\\
$6\,M_\odot$   & $1600\,L_\odot$ & $\ga 0.05\,M_\odot$\\
$10\,M_\odot$   & $10\,000\,L_\odot$ & $\ga 0.025\,M_\odot$\\
$20\,M_\odot$   & $170\,000\,L_\odot$ & $\ga 0.0125\,M_\odot$\\
\noalign{\smallskip}
\hline
\end{tabular}
\end{table}

The values for the minimum envelope masses drop strongly
with increasing protostellar mass and luminosity.
As the observed circumstellar masses for solar-mass Class~0 protostars in nearby star forming regions
are typically around $1\,M_\odot$ \citep[][]{submm-survey-low-mass-ps}, i.e.~a factor of 
two above our detection limit for $M_{\rm protostar} = 1\,M_\odot$ protostars,
we can expect to detect solar-mass protostars in the Carina Nebula
in our \textit{Herschel} maps. Less massive protostars 
($M_{\rm protostar} < 1\,M_\odot$) are
 generally not sufficiently
luminous to produce far-infrared fluxes above our detection limits.

\paragraph{YSOs with circumstellar disks (T~Tauri stars):}  

For more evolved YSOs, where much of the circumstellar material is
in a circumstellar disks (but significant envelopes may still also be
present), there are numerous possible spatial configurations
for the circumstellar material.
We therefore considered the  \cite{robitaille-grid} YSO models, that contain
YSOs with a wide range of different masses and evolutionary stages.
We first selected from the grid of 20,000 models all those that represent
YSOs with a specific stellar mass, and then determined which of these models
would produce sufficiently strong fluxes for a detection in our
\textit{Herschel} maps and what the circumstellar (i.e.~disk + envelope) mass
of these models is.
This analysis lead to the following results:

For $M_\ast = 1\,M_\odot$ YSOs, most models with circumstellar mass $\ga 0.5\,M_\odot$ are above our \textit{Herschel} detection limits; the lowest circumstellar mass of all detectable models is $\approx 0.1\,M_\odot$.
For $M_\ast = 3\,M_\odot$ YSOs, most models with circumstellar mass $ \ga 0.1\,M_\odot$
are above the detection limits; the lowest circumstellar mass of all detectable models is is $\approx 0.05\,M_\odot$.

For $M_\ast = 6\,M_\odot$ YSOs, most models with circumstellar mass $\ga 0.01\,M_\odot$
are above the detection limits; the lowest circumstellar mass of all detectable models is is $\approx 0.003\,M_\odot$.

For $M_\ast = 10\,M_\odot$ YSOs, all models with circumstellar mass $\ge 0.0025\,M_\odot$
are above the detection limits. 

To put these numbers in the proper context, we have to compare them
to the observed circumstellar mass of YSOs in nearby star forming regions.
For T~Tauri stars, i.e.~low-mass YSOs ($M_\ast \le 2\,M_\odot$) with ages between
$\sim 0.5$~Myr and a few Myr, 
disk masses of up to  $M_{\rm disk} \sim 0.2\,M_\odot$ have been determined for some 
objects \citep[][]{orion-disks}, but
the median disk masses are much lower, only around $M_{\rm disk} \sim 0.03\,M_\odot$ \citep[][]{proplyds-massive-disks-orion}.
This implies that T~Tauri stars with typical disk masses remain undetectable 
in our \textit{Herschel} maps; we can only expect to detect a small fraction of the 
youngest T~Tauri stars with particularly massive disks.

For intermediate-mass ($M_\ast \approx 2 - 10\,M_\odot$) YSOs, i.e.~objects in the 
regime of Herbig Ae/Be stars, typical disk masses
are about $ 0.03\,M_\odot$ \citep[][]{disks-around-myso-herbig-ae,gas-dust-evolution-herbig-ae}.
Herbig Ae/Be stars with masses of at least $M_\ast \sim 4\,M_\odot$ are sufficiently luminous to 
be generally detectable in our \textit{Herschel} maps.

\subsubsection{Conclusions on the detection limit for YSOs}

From these limits it is clear that we can detect only a small fraction of all
YSOs in the Carina Nebula:
YSOs with (proto-)~stellar masses below $\approx 1\,M_\odot$
are usually undetectable (unless they would have exceptionally massive 
disks or envelopes).

According to the 
model representation of the field star IMF by \citet{kroupa-imf-uniform-in-var-sys},
the number of stars with masses below $1\,M_\odot$
is about 10 times larger than the number of stars with masses above $1\,M_\odot$.
This implies that we can detect only a few percent of the total 
young stellar and protostellar
population as point-like sources in our \textit{Herschel} maps.

\subsection{Contamination}

Although most of the far-infrared sources seen in our
\textit{Herschel} maps of the CNC will
be YSOs, there may be some level of contamination by other kinds
of objects.
The two most relevant classes of possible contaminants are
evolved stars and extragalactic objects.

\subsubsection{Evolved stars}

The source $\eta$~Carinae, which is a well known evolved massive star is of course excluded from our sample of YSOs. This object will be discussed in detail in Sec.~\ref{sec:The far-infrared spectral energy distribution of Eta Car}.

Evolved stars experience high mass loss and are often surrounded by
dusty circumstellar envelopes that produce strong excess emission
at far-infrared wavelengths \citep[e.g.][]{agb-laboca}.
The infrared SEDs of evolved stars can be quite similar to those of
YSOs, and thus the nature of objects selected by criteria based on
infrared excess alone is not immediately clear and can lead to ambiguities \citep[e.g.][]{post-agb-vs-pms}.\\
In our case, the location of the vast majority of far-infrared sources inside (or close to) the molecular clouds clearly suggests that they are most likely YSOs.
However, there is always the possibility that an unrelated
background source \textit{behind} the cloud complex may just appear
to be located \textit{in} the clouds.

As a first check
for possible contamination of our sample by evolved stars,
we used the SIMBAD database\footnote{http://simbad.u-strasbg.fr/simbad/} to search for AGB, SG, and RGB type stars within a $5\arcsec$ radius around each of our 631~\textit{Herschel} point-like sources.
The SIMBAD database lists 82 objects of the above mentioned type of object within
the area of our maps, but none of these is associated with one of 
our \textit{Herschel} point-like sources,
suggesting that these known evolved stars in our field-of-view are 
too faint at far-infrared wavelengths
to be detected in our \textit{Herschel} maps.

\smallskip

In a further  approach to quantify the possible level of contamination by evolved stars,
we  used the models of \cite{agb-evolution} to compute the expected far-infrared fluxes of asymptotic giant branch stars.
Using the online tool\footnote{http://stev.oapd.inaf.it/cgi-bin/cmd\_2.3} we computed
10\,357 stellar models\footnote{The following parameters were used:
Metallicity: $Z=0.02$; Age: $\log(t/{\rm yr})=6.6-10.13$ at steps of $\log(t/{\rm yr})=0.05$;
dust composition as in \cite{agb-ir-colors}: 60\% Silicate + 40\% AlOx for M stars and 85\% AMC + 15\% SiC for
C stars; Chabrier (2001) lognormal IMF for single stars}.
We then determined the far-infrared fluxes at the  wavelengths of $60\,\mu$m and $160\,\mu$m
for a model  distance of $d\ge2.5~\rm{kpc}$ (i.e., assuming that these evolved stars were located
immediately behind the Carina Nebula).
We found that only $\le147$ of the model AGB stars (i.e.~$\le1.4\%$ of the total sample)
would have far-infrared fluxes above our detection limits of $\sim 1$~Jy.
This suggests that the probability to detect evolved stars located in the Galactic background
behind the Carina Nebula is low.\\
\cite{agb-mass-loss-herschel} observed $\sim150$ evolved stars with \textit{Herschel}. They found the typical mean flux ratio of the PACS $70\,\mu$m and the PACS $160\,\mu$m flux for such stars to be $\approx6.1\pm1.5$. One can see that all but four of our sources have 160/70 ratios larger than this value, which is another indicator that the contamination of our sample of \textit{Herschel} sources with evolved stars is very small.

\subsubsection{Extragalactic contaminants}

The possible level of extragalactic contamination can be
determined from the results of \cite{herschel-extragal-contamin}
who give galaxy number counts obtained
from SPIRE observations for the first 14~deg$^2$ of the Herschel-ATLAS survey \citep[][]{herschel-atlas}.
The highest measured flux of any galaxy in this sample was 0.8~Jy, 
i.e.~below our detection limit. Therefore, it appears very unlikely that our sample of point-like
 \textit{Herschel} sources in the Carina Nebula contains extragalactic objects.

In conclusion, we find that the possible level of contamination must be very low.
Most likely, all our \textit{Herschel} point-like sources are YSOs associated to the Carina Nebula and will from now on be called YSO candidates.

\section{Modeling the spectral energy distributions of the \textit{Herschel} detected YSOs}\label{sec:Analysis of the point-like sources}

In order to derive information about the properties of the \textit{Herschel} detected YSOs,
we assembled their spectral energy distributions (SEDs) over an as wide as possible wavelength range
and compared them to radiative transfer models. For this analysis we focused on the Carina Nebula (see Fig.~\ref{img:analyzed-region-cut}). The 67 objects associated with the distant molecular clouds at the eastern and western edge of our \textit{Herschel} maps were excluded.
Furthermore, the 92 sources in the region of the Gum~31 cloud will be analyzed
in a separate study (Ohlendorf et al. 2012, A\&A submitted). This leaves us with a total number of 482 \textit{Herschel} point-like sources with fluxes detected in at least two bands.

Since the reliability of SED modeling depends strongly on the wavelength coverage,
we considered only those objects from
our \textit{Herschel} catalog of point-like sources (see Sec.~\ref{sec:The Herschel point source catalog}) 
that were detected in at least three \textit{Herschel} bands and at least one \textit{Spitzer} IRAC band.
We performed a careful visual inspection of the  \textit{Spitzer} IRAC images to make sure
that only those \textit{Herschel} point-like sources that can be clearly identified by an
apparently single \textit{Spitzer} counterpart were included in the sample. Many \textit{Herschel} point-like sources turned out to have no, unclear, or multiple \textit{Spitzer} counterparts; these were rejected from the sample.  As a further check, we also inspected 
our deep near-infrared VLT HAWK-I images \citep[][]{carina-hawki} for those \textit{Herschel} 
sources that are located in the HAWK-I field-of-view and excluded two \textit{Herschel} sources
that turned out to be very compact star clusters.
This procedure left us with a final sample of 80 reliable apparently single point-like sources 
with now fluxes in at least three \textit{Herschel} bands and at least one \textit{Spitzer} band. 
For 36 of these, we found apparently single counterparts in the 
\textit{2MASS} near-infrared images and added their
near-infrared magnitudes as listed in the \textit{2MASS} point source catalog 
\citep[][]{2mass} to the SED. Apparently single counterparts in the \textit{WISE} All-Sky Data Release Catalog (Cutri et al. 2012) were found for 38 of these 80 \textit{Herschel} sources and the 12 and $22\,\mu$m photometry was added. The complete sample of the 80 sources with all available fluxes used for the SED fitting can be found in Table~\ref{tbl:phot-fluxes-201}.\\
The number of \textit{Herschel} point-like sources in the CNC without a clear \textit{Spitzer} counterpart, but at least three \textit{Herschel} fluxes, is 241.

\subsection{SED fitting with the Robitaille models}\label{sec:The Robitaille models for young stellar objects}

\cite{robitaille-grid} present a grid of 20,000 models\footnote{All models are publicly available at \emph{http://caravan.astro.wisc.edu/protostars/}}
 of young stellar objects 
(YSOs) which were computed using a 2D radiative transfer code developed by 
\cite{whitney-rad}. These models describe YSOs with a wide range of masses
and in different evolutionary stages, from the early stages of protostars
embedded in dense in-falling envelope,
 until the late pre-main sequence stage, when only a remnant disk is left. 
These models are characterized by numerous parameters
describing the properties of the central object (e.g.~mass, luminosity, 
temperature), the circumstellar envelope (e.g.~outer radius, 
envelope accretion rate, opening angle of a cavity),  and
the circumstellar disk (e.g.~mass, outer disk radius, disk accretion rate, 
flaring). For each model, SEDs are given
 for ten different inclinations, resulting in a total of 200,000 model SEDs.

To fit our sample of point-like sources, an IDL routine was implemented from the code of \cite{robitaille-sed} and \cite{robitaille-grid}. The 80 point-like sources that have been analyzed have up to 14 fluxes from the five \textit{Herschel} bands, the four \textit{Spitzer} IRAC bands, the three \textit{2MASS} bands, and the \textit{WISE} 12 and $22\,\mu$m bands. The distance was fixed to 2.3~kpc. The interstellar extinction was restricted to the range of $A_{\rm v} = [0...40]$ mag. Finally, an error of 10\% was assigned to the \textit{2MASS} fluxes, an error of 20\% to the \textit{Spitzer} fluxes, and an error of 30\% to the \textit{Herschel} and \textit{WISE} fluxes\footnote{For none of our sources the formal photometric uncertainties for a given flux measurement exceeded these assigned error values.}

 \begin{figure}[!htb]
 \centering
    \includegraphics[width=8.8cm, keepaspectratio]{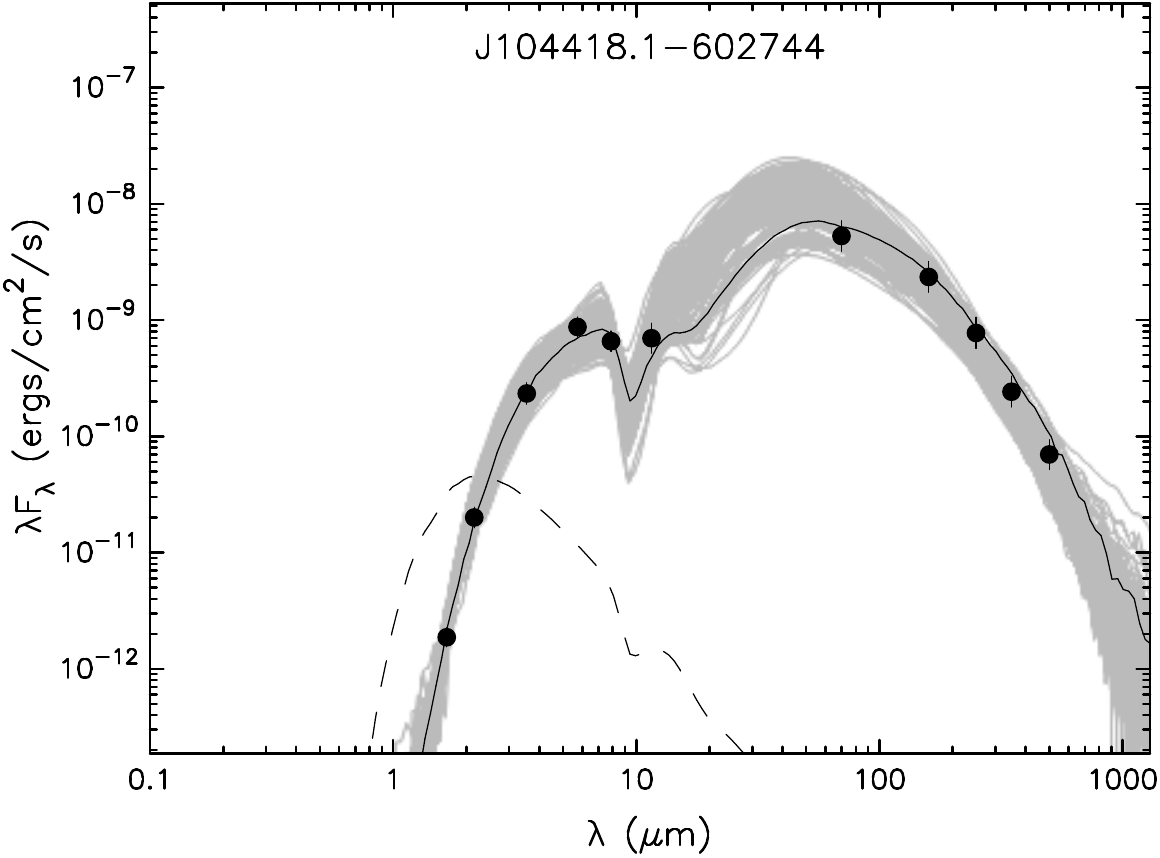}
    \caption{Example of a IR source with a good SED fit. The filled circles show the fluxes. The black line shows the best fit, and the gray lines show subsequent good fits with $\chi_{\nu}^2-\chi_{\nu, \rm best}^2<2$. The dashed line shows the stellar photosphere corresponding to the central source of the best fitting model, as it would look in the absence of circumstellar dust (but including interstellar extinction).}
     \label{img:sed-vergleich}%
 \end{figure}

\subsection{Stellar and circumstellar parameters of the YSO candidates}\label{sec:Parameters of the protostar candidates obtained from SED fitting}

\begin{figure*}
\centering
  \includegraphics[width=\linewidth, height=\textheight, keepaspectratio]{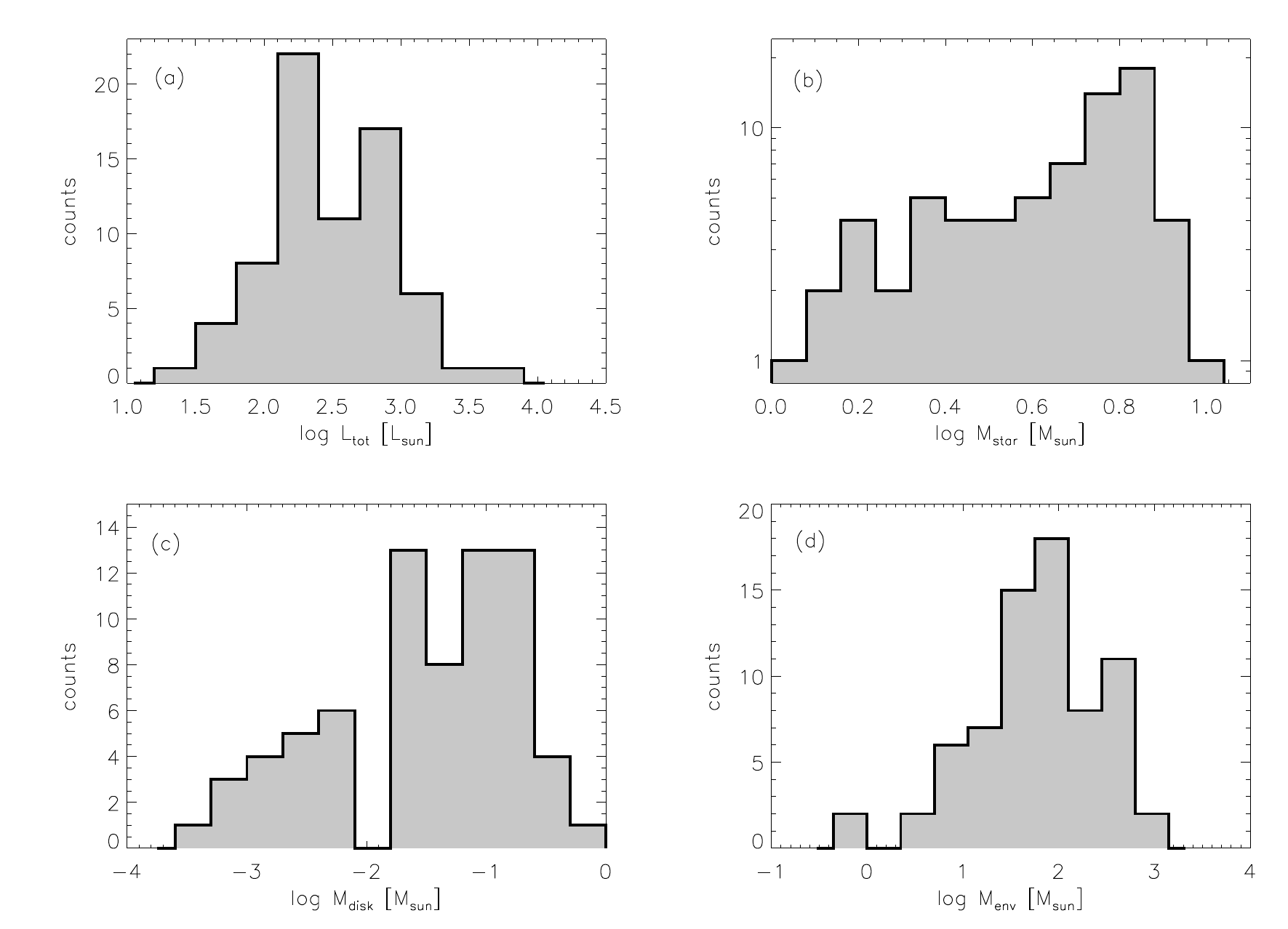} 
\caption{Histograms of the four model parameters obtained from the SED analysis for sources with an acceptable SED fit: Total luminosity (a), central stellar mass (b), circumstellar disk mass (c), and envelope mass (d).}
\label{img:model-histo}
\end{figure*}

For 71 of the 80 sources, acceptable fits ($\chi_{\nu}^2<5$)\footnote{$\chi_{\nu}^2=\chi^2$ per data point. Note that this is not the formal statistical definition of a reduced $\chi^2$.} were found. An example of such an acceptable fit is presented in Fig.~\ref{img:sed-vergleich} for source J104418.1$-$602744.
Before considering the resulting best-fit values of the stellar and circumstellar parameters,
we emphasize the well known fact that the results of SED fits can be
highly ambiguous \citep[e.g.][]{accuracy-protostars-sed}.
Many of the stellar and circumstellar
parameters are often poorly constrained because the models show a high degree of degeneracy \citep[e.g.][]{radiation-transfer-circumstellar-disk}. We therefore restrict our analysis to a few selected parameters that can be
relatively well determined from these fits. These are the total luminosity, the
stellar mass, and the mass of the circumstellar envelope.
Histograms for these three model parameters, as well as for the circumstellar disk mass, can be found in Fig.~\ref{img:model-histo}; the
values and their individual uncertainties are listed in Table~\ref{tbl:model-param-ranges}.

\begin{figure*}
\centering
  \includegraphics[width=\linewidth, height=\textheight, keepaspectratio]{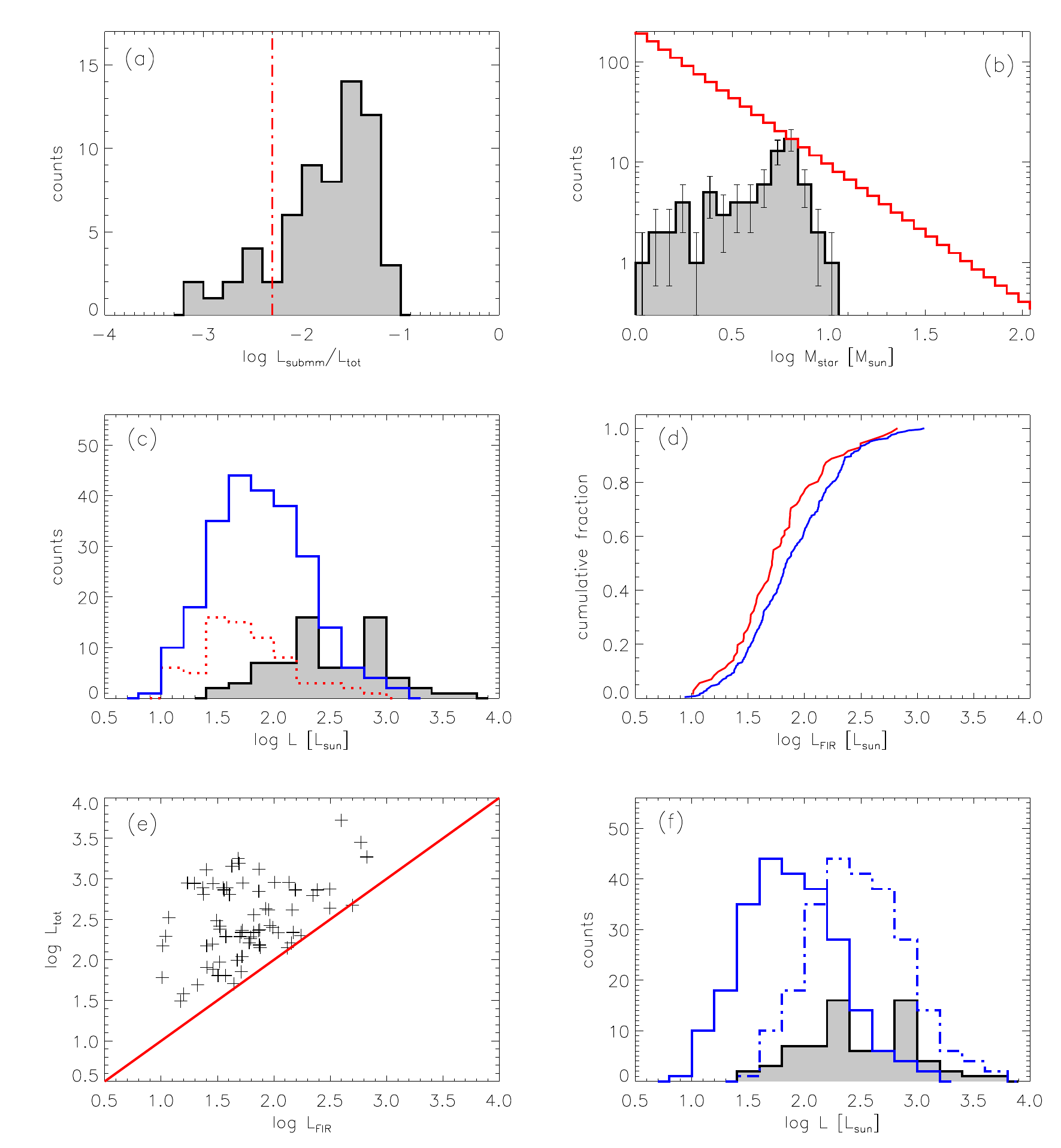} 
\caption{Total luminosities and central stellar mass of the \textit{Herschel} YSOs. \textit{Top panels:} (a) Ratio of the sub-mm luminosity (integrated SED for $\lambda\ge350\,\mu$m) and the total luminosity obtained from the Robitaille models for all sources with an acceptable SED fit. The vertical dash-dotted red line marks the transition between Class~0 ($L_{\rm submm}/L_{\rm tot}\ge0.005$) and Class~I objects \citep[see][]{from-pre-to-protostar-andre1993}. (b) Histogram of the stellar mass. The solid red line shows the IMF with a slope of $\gamma = -1.35$. 
\textit{Middle panels:} (c) Histogram of the total luminosity obtained from the Robitaille models for sources with an acceptable SED fit (gray histogram). Overploted are the integrated far-infrared luminosity $L_{\rm int,FIR}$ of all sources with an acceptable SED fit (dotted red histogram), and the integrated far-infrared luminosity of all \textit{Herschel} point-like sources in the CNC with fluxes detected in minimum
 three bands and without an SED fit (solid blue histogram). (d) Cumulative distribution function of the
 integrated far-infrared luminosity of all sources with an acceptable SED fit (red pluses), and the
 integrated far-infrared luminosity of all \textit{Herschel} point-like sources in the CNC with fluxes
 detected in minimum three bands and without an SED fit (blue crosses). \textit{Bottom panels:} (e) Total
 luminosity obtained from the Robitaille models versus the integrated far-infrared luminosity for all
 sources with an acceptable SED fit. The solid red line marks a ratio of 1. (f) Same as (c), but now with
 the distribution of the integrated far-infrared luminosity of all \textit{Herschel} point-like sources in
 the CNC with fluxes detected in minimum three bands and without an SED fit (solid blue) multiplied by the
 median value of $L_{tot}/L_{FIR}$ for YSOs with SED fit (dash-dotted blue).}
\label{img:model-histo2}
\end{figure*}

\paragraph{Total luminosity:}
The  total luminosity is relatively well determined by the amplitude of the SED, because
the \textit{Herschel} bands cover the broad far-infrared peak of the SED quite well.
The total luminosities of the YSOs derived from the fits range from $\approx 30\,L_\odot$ to
$\approx 5300\,L_\odot$. The lower boundary is a result of the detection limit.
The rather moderate value of the upper boundary, however, will lead to interesting 
conclusions about the currently forming stellar population, as will be discussed
below.

\paragraph{(Proto-)~stellar mass:}
The (proto-)~stellar masses are rather tightly related to the total
luminosities in these models.
The derived values range from $1\,M_\odot$ up to $\approx 10\,M_\odot$. 
The lower boundary is again the  result of the detection limit and agrees well
with the estimates discussed above.
The upper boundary, however, is surprisingly low, given the fact that the Carina Nebula
contains at least 70 stars with masses of well above $20\,M_\odot$, including
numerous very massive ($M \ga 50\,M_\odot$) stars.
The derived mass distribution for the \textit{Herschel} detected YSOs suggests that the currently 
forming generation of stars in the CNC is restricted to intermediate- and low-mass stars,
but does not seem to form stars as massive as present in large numbers in the slightly
older population of optically visible stars.
We note that the lack of high-mass YSOs is \textit{not} an artifact of the
model grid: the Robitaille grid contains objects with masses up to $50\,M_\odot$.
A more detailed discussion of these aspects will be presented in Sec.~\ref{sec:yso-mass-function}.

\paragraph{Circumstellar disk mass:}
The disk mass is not a very well constrained parameter, because there are large
ambiguities with the envelope mass. Nevertheless, we note that the best-fit values
range between $0.0004\,M_\odot$ and $0.6\,M_\odot$.

\paragraph{Envelope mass:}
The rather high values we find for the envelope masses (between $10\,M_\odot$ and $1000\,M_\odot$)
confirm the expectation that most of the \textit{Herschel} detected YSOs are protostars
still embedded in rather massive envelopes.

\subsection{Sub-mm luminosities of the point-like sources}\label{sec:Sub-mm luminosities of the point-like sources}

The observational definition of protostars
is based on the ratio of the sub-mm luminosity to the
total luminosity of a YSO.
Objects with fractional sub-mm luminosity of at least
$L_{\rm submm} / L_{\rm tot} \ge 0.005$ are defined as ``Class~0 protostars'' \citep[][]{from-pre-to-protostar-andre1993}. Since this ratio requires knowledge of the total luminosity, we can only apply 
it to the YSO in the SED fit sample.

We calculated the sub-mm luminosity for all sources with an SED fit, 
integrating the SED for $\lambda\ge350\,\mu$m, which in this case resulted in 
the sum over the two longer SPIRE bands at 350 and $500\,\mu$m:
$ L_{\rm submm}=F_{350}\times {\rm d}\nu_{350} \, + \, F_{500}\times {\rm d}\nu_{500}$ 
with $F_{\nu}$ the flux density and ${\rm d}\nu$ the width of the band filter.
The resulting values are shown as a histogram in Fig.~\ref{img:model-histo2}a. 
Fifty-three of the 71 objects, i.e.~75\%,  can be classified 
as Class~0 protostars.

The fraction of Class~0 protostars among the 402 \textit{Herschel} YSOs
for which no clear \textit{Spitzer} counterparts could be found
is most likely even larger (because the absence of a \textit{Spitzer}
counterpart suggests the object to be a pre-stellar core or a very
young protostar).
This demonstrates that the sample of \textit{Herschel}-detected YSOs
traces the extremely young population of currently forming stars;
these objects are systematically younger than the YSO population
revealed by the \textit{Spitzer} observations \citep[][]{carina-south-pillars-spitzer,carina-south-pillars-spitzer-protostars-intermediate}.

\subsection{Luminosities of the YSOs without SED fit}

The 71 objects with acceptable SED fits represent only 15\% of the 
total number of \textit{Herschel} point-like sources in the analyzed
area of the Carina Nebula.
For most of the remaining sources, no clear counterparts at shorter wavelengths
could be found. Some of these objects may be pre-stellar cores, but
some could also be protostars with very dense envelopes that prevent
their detection in the \textit{Spitzer} maps.

The only information that can be derived about these objects from the
present data is the far-infrared luminosity,
 integrated over the wavelength range covered by our \textit{Herschel} data.
We computed these values as the sum of the observed fluxes multiplied by the bandwidth. The resulting far-infrared luminosities are shown in the histogram in Fig.~\ref{img:model-histo2}c. The median of this distribution is $69.2\,L_\odot$ and the maximum is $1150\,L_\odot$.

In order to investigate how similar or different the \textit{Herschel}-detected
objects without clear \textit{Spitzer} counterparts
are with respect of those that have clear \textit{Spitzer} counterparts,
we compare the far-infrared luminosities of these two groups in Fig.~\ref{img:model-histo2}c.
Their cumulative distribution functions are shown in Fig.~\ref{img:model-histo2}d.
A Kolmogorov-Smirnov test gives a probability of $P_0=0.08$ that both samples are drawn from
the same parent distribution. This indicates that the far-infrared luminosities of the sources with a clear \textit{Spitzer} counterpart are slightly systematically lower (by about 30\%) than those without \textit{Spitzer} counterpart. However, the statistical significance of this difference is marginal, and we thus can assume that also the distribution of total luminosities in the full sample of \textit{Herschel}-detected objects should be similar
to those of the objects with SED fit.

In Fig.~\ref{img:model-histo2}e we compare the far-infrared luminosities of the objects
with SED fit to their total luminosities. The median value for the ratio
$L_{\rm tot} / L_{\rm FIR}$ of this sample is 4.25.
Multiplying the distribution of far-infrared luminosities of the objects without SED fit
by this factor
can thus give us a (crude) estimate of the distribution of their total luminosities.
The resulting distribution, based on this simple extrapolation factor, is shown
in Fig.~\ref{img:model-histo2}f.
One can see that the extrapolated distribution of total luminosities agrees
reasonably well with the  distribution of total luminosities for the objects
with SED fits. The most important point is that the extrapolated total luminosities
are again restricted to values below $\la 5000\,L_\odot$.\\
At this point we note that \textit{Herschel} sources without \textit{Spitzer} detections are expected to have a systematically lower $L_{\rm tot} / L_{\rm FIR}$ than sources with \textit{Spitzer} detections.
Although there are substantial uncertainties in this extrapolation, these results
suggest that there is no significant number of YSO with total luminosities
exceeding $10\,000\,L_\odot$, i.e.~the lower limit for high-mass YSOs.

\section{The mass function of the Herschel detected YSOs}\label{sec:yso-mass-function}

\subsection{The apparent deficit of massive YSOs}

Our SED modeling suggests that all of the analyzed \textit{Herschel} 
YSOs in the CNC are YSOs of low- or intermediate mass,
$M_\ast \la 10\,M_\odot$. 
The result of a lack of high-mass YSOs is also corroborated
by the fact that none of the \textit{Herschel} detected YSOs in our full sample
has a luminosity of more than $10^4\,L_\odot$  (which is the
lower boundary for high-mass YSOs). If a massive YSO with such a high
luminosity existed in the Carina Nebula, there is no reason
why it should not be detected as a very prominent and bright
far-infrared source in our \textit{Herschel} maps.

The absence of massive YSOs is also supported by the lack of hyper-and ultra-compact H\,II regions. This is quite remarkable, given the large number of high-mass stars
in the young stellar populations in the Carina Nebula:
the compilation of \cite{carina-o-stars} lists 70 O-type stars
(with stellar masses $\ge 20\,M_\odot$), among which
there are 18 stars with stellar masses $\ge 50\,M_\odot$.

To illustrate the lack of massive stars among the
\textit{Herschel} detected YSOs, we compare in Fig.~\ref{img:model-histo2}b their
mass distribution to the shape of the field-star
IMF\footnote{We note that there is no evidence that the
IMF in the Carina Nebula would deviate from the canonical
field star IMF.}.
For masses above $\approx 6\,M_\odot$, the observed
distribution of YSO masses 
drops much more quickly with increasing mass than the
field star IMF and reveals an apparent deficit of stars $M_* \ga 10\,M_\odot$.

\subsection{Detection limits and biases}

A similar result was obtained by \cite{carina-south-pillars-spitzer-protostars-intermediate} from their analysis
of their \textit{Spitzer} selected YSO sample: they also found
no YSO with masses above $M_\ast\approx10\,M_\odot$.
They interpreted this as an effect of the infrared-excess selection
of their sample:
since massive stars disperse their disks on shorter timescales 
than lower mass stars, they display infrared excesses (and thus are
detectable via infrared excess emission) for a shorter
period of time. The  ``canonical'' disk lifetime for 
solar-mass YSOs is  $\sim 2-3$~Myr \citep[see][]{ic348-spitzer-census,pms-timescales-accretion},
i.e.~considerably longer than the $0.1 -1 $~Myr disk lifetime
determined for intermediate-mass YSOs \citep[][]{herbig-ae-in-ob-associations,carina-south-pillars-spitzer-protostars-intermediate,disc-evolution-in-ob-ass,herbig-ae-submm-survey}.
As the \textit{Spitzer} data are sensitive enough
to detect several Myr old solar-mass YSOs with rather low disk masses,
this is a valid explanation for the lack of massive objects in the 
\textit{Spitzer} excess-selected YSO sample.

For our \textit{Herschel} selected sample, the situation is 
different, because the detection limits are much more restrictive.
The relevant timescale is \textit{not} the disk lifetime,
but the timescale, for which a YSO still has a sufficient amount of
circumstellar material to produce enough far-infrared emission to be detectable.
As discussed above in Sec.~\ref{sec:Herschel sensitivity to circumstellar matter}, solar-mass YSOs can only be detected 
in our \textit{Herschel} maps 
during their early protostellar phase 
or as long as they have exceptionally massive disks.
The typical timescale for which such solar-mass YSOs are detectable
is thus the duration of the protostellar phase, i.e.~about $0.1$~Myr
\citep[see][]{low-mass-sf-observations-evans}
which is much shorter than the ``canonical'' disk lifetime.
Another important aspect is the strong
dependence of the minimum required circumstellar mass for a
\textit{Herschel} detection on the luminosity (and thus the mass)
of the YSO. As determined above, the minimal required circumstellar mass 
decreases from $\sim 0.5\,M_\odot$ for $M_\ast = 1\,M_\odot$ YSOs, via
$\sim 0.01 - 0.05\,M_\odot$ for $M_\ast = 6\,M_\odot$ YSOs to 
$\sim 0.002 - 0.01\,M_\odot$ for $M_\ast = 20\,M_\odot$ YSOs.
This leads to a situation where the period of time, during which
the YSOs are detectable for \textit{Herschel}, is \textit{not} 
decreasing with increasing stellar mass.

For high-mass ($M \ge 10\,M_\odot$) stars, 
the lifetime of circumstellar material is not well known,
since the details of the formation mechanism of high-mass stars
are still not well understood \citep[][]{high-mass-sf-review}.
From an observational point of view, protostars of
$\leq 20\,M_\odot$ are often surrounded by disks containing several
solar masses of circumstellar matter \citep[e.g.][]{disk-dust-around-high-mass-ps,disks-around-young-ob}. For higher protostellar masses, the situation is still unclear since
no good examples of proto O-stars  have been found so far.
We thus have to consider the results of numerical simulations
of massive star formation.
The calculations of \cite{massive-star-formation-simulation}, \cite{massive-star-formation-accretion}, \cite{simulation-massive-sf-disk-accretion}, and \cite{radiative-feedback-cluster} showed that the forming massive stars
are surrounded by considerable amounts of circumstellar matter
(of the order of a few solar-masses, i.e.~well
above the minimum circumstellar mass required for
a \textit{Herschel} detection)
for at least about 50\,000 -- 100\,000~years.
This suggests that the period of time during which YSOs are
detectable for \textit{Herschel} is not a strong function of 
YSO mass. This agrees and is supported by the estimate of $\approx10^5$~yr for massive YSO lifetimes by \cite{lumi-and-timescale-myso}.

\subsection{Quantification of the massive YSO deficit}\label{sec:Quantification of the massive YSO deficit}

In order to quantify the deficit of massive YSO, we consider
the number of YSOs in the $[5 - 7]\,M_\odot$ mass range.
This range covers the peak of the observed YSO mass function,
and its lower end is high enough not be affected by incompleteness
of detection.
Our sample of \textit{Herschel}-detected YSOs with acceptable SED fits
contains 31 objects in the 
$[5 - 7]\,M_\odot$ mass range.

According to the model representation of the field star IMF by \citet{kroupa-imf-uniform-in-var-sys}, the ratio of the number of stars in the $[10 - 100]\,M_\odot$ mass range
to those in the  $[5 - 7]\,M_\odot$ mass range is 1.09.

Therefore, assuming a field-star IMF, the expected number of YSOs
in the  $[10 - 100]\,M_\odot$ mass range would be $\approx 34$.\\
Even if we (conservatively) assume that the period of time during which 
the massive YSO are detectable for \textit{Herschel} is a factor of three
shorter than those of the $[5 - 7]\,M_\odot$ YSOs,
the expected number of high-mass YSOs would be about 11, whereas the
actually observed number is zero.

It is highly unlikely that the non-detection of such objects
 is a statistical effect, since
the Poisson probability to detect no object if the expectation value is 11, is $e^{-11} = 1.7 \times 10^{-5}$.

We can thus conclude that the mass distribution of the
currently forming generation of stars detected by \textit{Herschel}
is different from the IMF of the optically visible population
of the  several Myr old stellar population in the Carina Nebula.
This difference seems to be related to the fact that nearly all the
clouds in which star formation is currently proceeding 
have too low densities and masses to allow the formation
of very massive stars \citep[see discussion in][]{carina-laboca}.

\section{Estimates for the size of the protostellar population and the star formation rate}

Since our \textit{Herschel} maps cover the full spatial extent of the Carina Nebula Complex,
the results can be used to estimate the total size of the protostellar population.
Considering the Carina Nebula Complex (including the area around Gum~31), but excluding the
67 objects in the distant molecular clouds at the eastern and western edge of our maps,
the total number of YSOs detected as \textit{Herschel} point-like sources is 574. Applying the
 criteria from \cite{herschel-epos-pre-high-mass-sf} for the distinction of pre- and protostellar cores, we
 consider these 267 \textit{Herschel} point-like sources with a $70\,\mu$m detection to be YSO (whereas the
 \textit{Herschel} sources without a 70 m detection may be pre-stellar cores). As
determined in Sect.~\ref{sec:Sub-mm luminosities of the point-like sources}, we can further 
assume that about 75\% of these 267 Herschel-detected YSOs are Class~0 protostars.
Hence the number of Herschel-detected protostars in
the entire CNC (including the Gum~31 region)  is 200.
To estimate the total number of protostars, we need an estimate of
the completeness of our sample. For this, we use the
modal value of the $70\,\mu$m flux distribution ($\approx 6$~Jy; see Fig.~\ref{img:herschel-flux-histos})
as an approximation of the detection completeness. In order
to find the corresponding protostellar mass limit, we again consider
the Robitaille models. We first selected from this grid models
representing YSOs with a specific stellar mass and additionally
fulfill the condition that their circumstellar (i.e.~disk+envelope) mass
is at least half of the stellar mass (this restricts the selection to protostellar objects).
Then we determined the stellar mass for which at least 50\% of the models in these
samples show $70\,\mu$m fluxes above the 6~Jy limit.
The resulting estimate of the completeness limit is about $2\,M_\odot$.
Assuming a Kroupa IMF, the number of stars in the $0.1-2\,M_\odot$ range
is approximately 20 times larger than the number of stars above $2\,M_\odot$.
Since the  number of \textit{Herschel}-detected protostars in the
CNC with $70\,\mu$m fluxes above the model value of 6~Jy is 144,
our estimate of the total protostellar population is
$\approx 144 \times 20 = 2880$.\\
If these protostars formed over a period of 100\,000 years
(i.e., the estimated lifetime of the protostellar phase), this
implies a star formation rate of about 0.029 stars per year.
Using the mean stellar mass of $0.6\,M_\odot$ (according to the
Kroupa field star IMF for the mass range $0.1-100\,M_\odot$), the
star formation rate of the CNC is then $0.017\,M_\odot/\rm{yr}$.\\
It is interesting to compare our result to the star formation determinations
by \cite{carina-south-pillars-spitzer-protostars-intermediate}. They derived a lower limit of $\ga0.008\,M_\odot/\rm{yr}$ for the recent star formation rate, averaged over the past 2~Myr, based on their analysis of a \textit{Spitzer}-selected sample of YSOs.  For the average star formation
rate over the past 5~Myr they derived $0.010-0.017\,M_\odot/\rm{yr}$.
For a meaningful comparison to our estimate, we have to take
into account that the area Povich considered is restricted to the
central 1.4 square-degrees of the Carina Nebula,
whereas our \textit{Herschel} sample covers the entire CNC, including the
Gum~31 region.
Considering these different areas, we find that 76\% of our \textit{Herschel}
point-like sources are in the area that also was studied by Povich.
Scaling our SFR estimate for the entire CNC by this factor, the
resulting rate of 0.013 for the central area agrees very well with
the rates determined by Povich.\\
We note that this good agreement of two completely independent estimates
is encouraging. It also suggests that the star formation activity in the
CNC remained approximately constant in the time from several Myr ago
until today.

\section{Spatial distribution of the YSO candidates}

\begin{figure*}[!htb]
\centering
   \includegraphics[width=\linewidth, keepaspectratio]{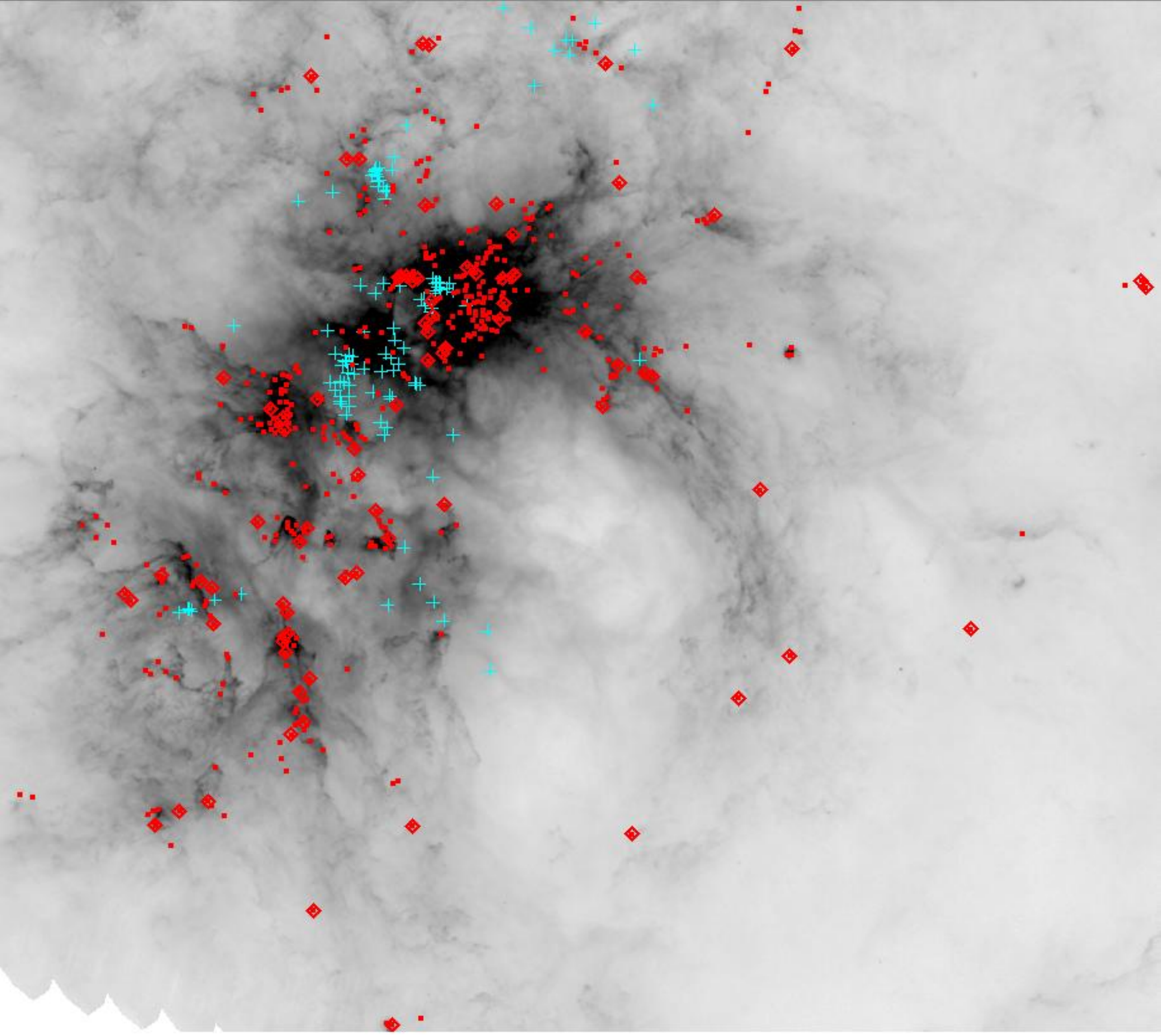}
   \caption{\textit{Herschel} PACS $70\,\mu$m image of the CNC with square root intensity scale. The positions of the OB stars from \cite{carina-o-stars} are marked with cyan crosses. Positions of the \textit{Herschel} point-like sources are marked with red squares. The positions of all 80 YSOs with SED fits are marked with red diamonds. Note that their distribution is concentrated on the cloud edges.}
    \label{img:carina-herschel-map-rgb-protostars}%
\end{figure*}

The spatial distribution of the YSO candidates is shown in
Fig.~\ref{img:carina-herschel-map-rgb-protostars}.
It is important to note that all clouds in the CNC are transparent at all
\textit{Herschel} wavelengths\footnote{As described in \cite{carina-herschel-clouds},
the column densities in 99\% of the area of our maps are
$N_{\rm H} \le 2.4 \times 10^{22}\,{\rm cm}^{-2}$ (corresponding to $A_{\rm v} \le 12$~mag)
and the corresponding optical depth in the $70\,\mu{\rm m}$ band is thus
$\tau(70\,\mu{\rm m}) \approx 0.036$.}.
This implies that cloud extinction is not an issue and
all (sufficiently luminous) YSOs should be detectable at all locations in our \textit{Herschel} maps.

Most \textit{Herschel} YSO candidates are located in the central regions of the Carina Nebula
and the South Pillars region.
In the northern part of the field, the source density is considerably lower.
A particularly interesting result is that the source density does \textit{not}
follow the distribution of cloud masses. The most prominent example for this
effect is the particularly massive and dense Northern Cloud (just to the west of the stellar cluster
Tr~14): although this cloud has a mass of about $50\,000\,M_\odot$ \citep{carina-herschel-clouds},
just about 30 YSO candidates are seen in the dense regions of the cloud.
Most of these are located at the eastern edge, where the cloud is
strongly irradiated by the numerous massive stars
in the Tr~16 and Tr~14 clusters.

Also in the other regions, the \textit{Herschel} YSOs are preferentially located at the surfaces of irradiated clouds or in narrow filaments.
This shows that the spatial distribution of the \textit{Herschel} YSOs (i.e.~the current star formation activity) does not follow the distribution of cloud mass,
but is largely restricted to locations of strong irradiation, i.e.~the edges of irradiated
clouds.

\begin{figure*}[!htb]
\centering
   \includegraphics[width=14cm, keepaspectratio]{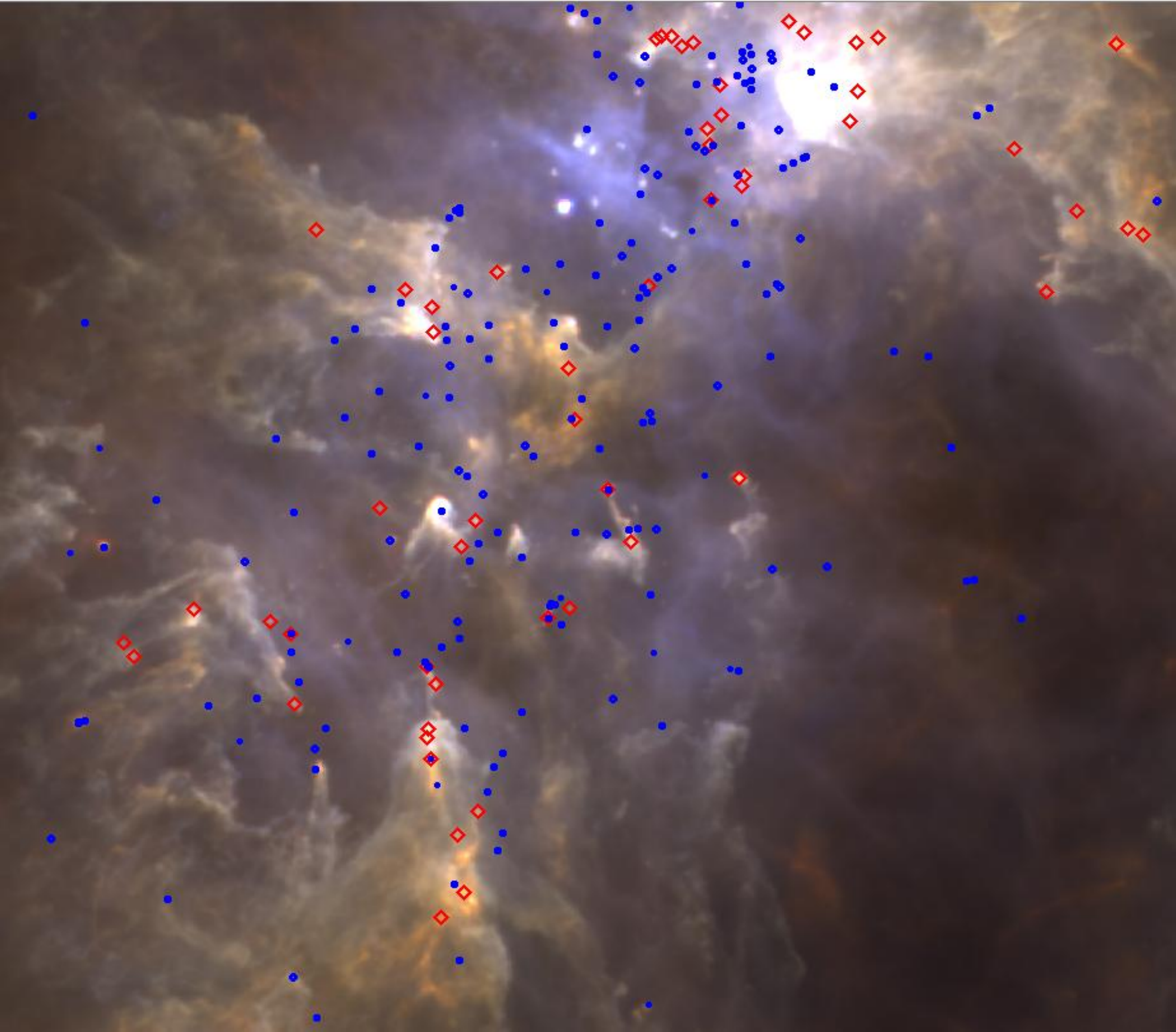}
   \caption[Same as Fig.~\ref{img:carina-herschel-map-rgb} with positions of YSO candidates.]
    {Same color composite image as in Fig.~\ref{img:analyzed-region-cut} of the South Pillars region. Positions of the YSO candidates are again marked with red diamonds. The blue circles represent the class~0 and class~I YSOs from the PCYC catalog by \cite{carina-south-pillars-spitzer-protostars-intermediate}. Note that the region around $\eta$~Car lacks \textit{Spitzer} YSOs because of the nebulosity caused by itself.}
   \label{img:carina-herschel-map-rgb-protostars-spitzer-ysos}%
\end{figure*}

In order to investigate this further, we show in
Fig.~\ref{img:carina-herschel-map-rgb-protostars-spitzer-ysos} a part of the central Carina Nebula and the Southern Pillars and compare the spatial distribution of the \textit{Herschel} YSO candidates
to the one of the \textit{Spitzer} YSO candidates from the Pan Carina YSO Catalog (PCYC) catalog \citep[][]{carina-south-pillars-spitzer-protostars-intermediate}.\\
These two samples should represent two different populations of young objects,
where the \textit{Herschel} sample is dominated by very young, deeply embedded
protostellar objects (with ages of about $\la 0.1$~Myr), while the PCYC sample should mostly consist of slightly older, more evolved young stars
(Class~I sources and T~Tauri stars).
Fig.~\ref{img:carina-herschel-map-rgb-protostars-spitzer-ysos} shows that most 
\textit{Herschel} YSO candidates are located
near the irradiated surfaces of clouds and pillars, whereas the \textit{Spitzer} selected YSO candidates often surround these pillars.

This characteristic spatial distribution of the young stellar populations
in different evolutionary stages agrees very well with the idea that the advancing
ionization fronts compress the clouds and lead to cloud collapse and star
formation in these clouds, just ahead of the ionization fronts. Some fraction of the cloud mass
is transformed into stars (and these are the YSOs detected by \textit{Herschel}),
while another fraction of the cloud material is dispersed by the process
of photo-evaporation. As time proceeds, the pillars shrink, and a population
of slightly older YSOs is left behind and revealed after the passage of the ionization front. This result
provides additional evidence that the formation of these YSOs was
indeed triggered by the advancing ionization fronts of the massive stars
as suggested by the theoretical models \citep[see][]{massive-stars-pillar-formation,carina-south-pillars-spitzer}.

\section{The far-infrared spectral energy distribution of $\eta$~Carinae}\label{sec:The far-infrared spectral energy distribution of Eta Car}
\begin{figure}[!htb]
\centering
\includegraphics[width=8.8cm, keepaspectratio]{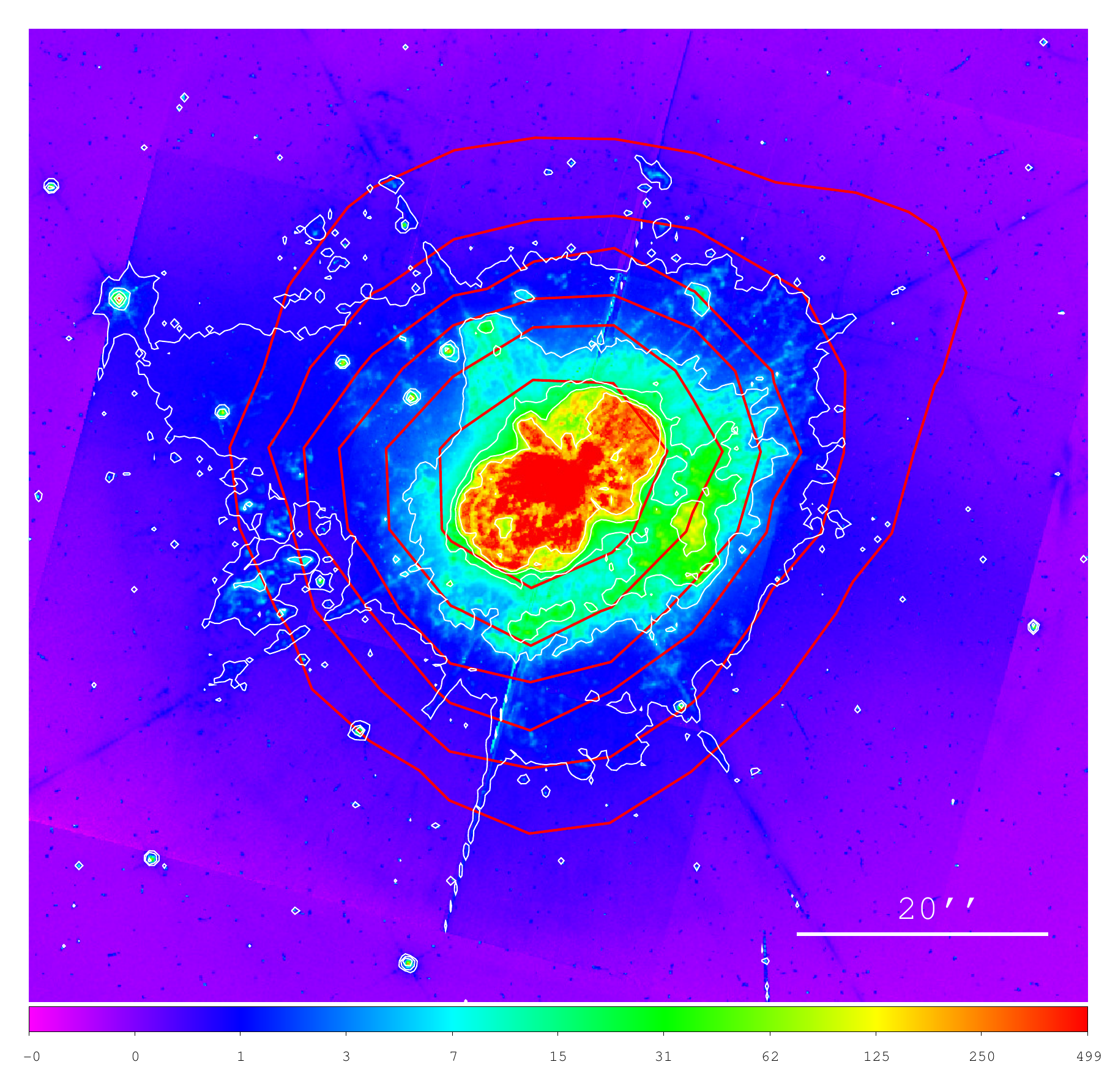}
\caption{HST/WFPC2 image (data set {\tt hst\_9226\_01\_wfpc2\_f658n\_wf}) of $\eta$~Carinae
obtained through the narrow-band filter F658N. The image is displayed with a logarithmic
intensity scale in order to show the rather faint nebular features of the
outer ejecta. The thin white contour lines also highlight the structure of the optical nebula.
The thick red contours trace the \textit{Herschel} $70\,\mu$m band emission.
The contour levels start at 3.125~Jy/pixel and increase by a factor of 2 up to the
200~Jy/pixel level.}
\label{img:eta-car-h70-hst}
\end{figure} 

The Luminous Blue Variable $\eta$~Carinae \citep[see][]{carina-eta-car-environments} is
one of the most luminous $(L_\ast \ge 5 \times 10^6\,L_\odot$) massive stars in our Galaxy.
Despite numerous observations in all wavelength regimes, the exact nature and evolutionary state
of this object remain as yet elusive \citep[][]{carina-eta-car-environments}.
The object seems to be a binary with a period of 5.5~years, and the extremely
strong stellar winds with a mass loss
rate of about $\dot{M} \sim 10^{-3}\,M_\odot/{\rm yr}$ cause very strong shocks and resulting
high-energy radiation from the wind-wind collision zone
\citep[see, e.g.][]{eta-car-material-wind-collision,eta-car-lhc}.
$\eta$~Car displays strong variability in almost all spectral regimes.
In the optical, it once represented the second brightest star on the sky, but faded
by more than eight magnitudes between 1850 and 1880. During the last three decades,
it brightened by several magnitudes
\citep[][]{eta-car-hst-chrysalis,carina-eta-car-revised-historical-lightcurve}. Strong X-ray variability is seen as a result of dynamical changes in the wind collision zone \citep[][]{eta-car-x-ray}.
The observed near-infrared variability
is probably related to the episodic formation of dust grains within compressed post-shock zones
of the colliding winds \citep[][]{eta-car-formation-winds}.

Our \textit{Herschel} images show a very bright and prominent compact source at the
position of $\eta$~Car.
The far-infrared emission originates from the circumstellar dust envelope
around $\eta$~Car.  The famous bipolar \textit{Homunculus Nebula}
is thought to be the result of the ``Great Eruption`` in the 1840's \citep[see][]{carina-eta-car-ir-morphology,
carina-eta-car-astrometry,carina-homunculus-hst-motion,carina-homunculus-review-smith,
carina-homunculus-nir-images,carina-eta-car-eruption-nature}.
The total dust + gas mass in the bipolar nebula and a dense equatorial torus in the
Homunculus is estimated to be about $15 - 20\,M_\odot$
\citep[][]{eta-car-torus-nature,eta-car-homunculus-mass-energy}. The angular diameter of the optically bright parts of the \textit{Homunculus Nebula} as seen in the
HST images is $18'' \times  11''$ (long axis $\times$ short axis).
The bright mid-infrared emission,  measured in an $18\,\mu$m image obtained
with the Magellan Telescope by \cite{eta-car-homunculus-mass-energy}, has the same extension.
This size scale implies that the \textit{Homunculus Nebula}
should be marginally resolved in our \textit{Herschel} PACS maps, but unresolved in the SPIRE maps.
However, the \textit{Homunculus Nebula} is surrounded by the so-called ``outer ejecta'',
a collection of numerous filaments, shaped irregularly and distributed over an area of $\approx 1' \times 1'$
\citep[][]{eta-car-ejecta-review}; the dust in these outer parts could also contribute to
the far-infrared emission.

In Fig.~\ref{img:eta-car-h70-hst} we show the contours of the \textit{Herschel} $70\,\mu$m emission
on an optical HST WFPC2 image taken through the narrow-band filter F658N
(data set ${\rm hst\_9226\_01\_wfpc2\_f658n\_wf}$, observed on 2001-06-04, exposure time 923.33 sec)
that shows the \textit{Homunculus Nebula} and the surrounding
outer ejecta.
A two dimensional Gaussian fit to the $70\,\mu$m emission in
(using the command {\tt Pick Object} in GAIA) yields a
FWHM size of $18.1'' \times 14.9''$, which is clearly larger
than the FWHM size of $10'' \times 10''$ measured for several isolated point-like sources
in the same map.
The measured direction of elongation is along a position angle of $33\degr$, well consistent
with the orientation of the  \textit{Homunculus Nebula}.

Due to the good angular resolution of our \textit{Herschel} maps, the compact far-infrared emission
around $\eta$~Carinae
can be well separated from the surrounding background.
We performed aperture photometry using circular apertures with radii of
$45\arcsec$ for the 70, 160, and $250\,\mu$m bands,
$50\arcsec$ for the $350\,\mu$m band, and $72\arcsec$ for the $500\,\mu$m band;
these apertures are large enough to include not only the
 \textit{Homunculus Nebula}, but also possible contributions from the
``outer ejecta'' (which extend up to radial distances of $30\arcsec$ from $\eta$~Car).
The fluxes derived in this way are
6685~Jy, 1163~Jy, 302~Jy, 138~Jy and 72~Jy,
for the 70, 160, 250, 350, and $500\,\mu$m band, respectively.

The resulting, very high, PACS fluxes have to be treated with caution,
since they are clearly in the non-linear regime ($\ge 200$~Jy) of the instrument.
The $70\,\mu$m flux, and also the $160\,\mu$m flux (although to a lower level),
is also affected by readout saturation, which starts at
200~Jy for PACS $70\,\mu$m and 1125~Jy for PACS $160\,\mu$m.
Since there is no experience with possible corrections for non-linearity and saturation at
such high flux levels (Herschel Helpdesk, priv. comm.), the
obtained PACS fluxes can only be used as lower limits to the
true fluxes. For SPIRE, the much lower measured source fluxes are below the instrumental
saturation level (Herschel Helpdesk, priv. comm.). As an additional check, we inspected
the raw data (timelines and masks),
but found no indications for truncations because of ADC saturation.
We thus can assume the derived SPIRE fluxes of $\eta$~Car to be reliable.

\begin{figure}[!htb]
\centering
\includegraphics[width=\linewidth, keepaspectratio]{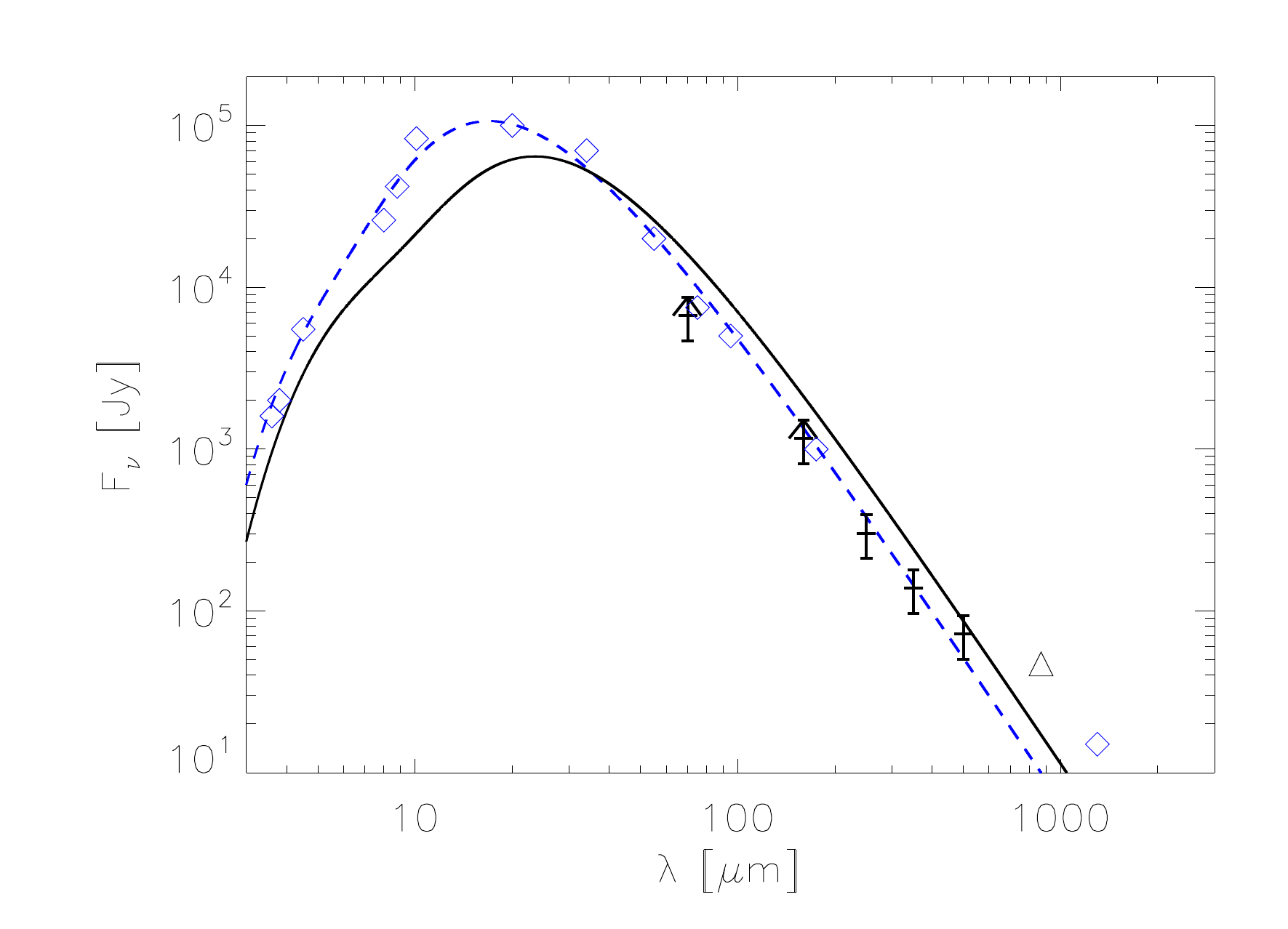}
\caption{Spectral Energy Distribution of $\eta$~Carinae. The black crosses show our \textit{Herschel} fluxes. The $70\,\mu$m and $160\,\mu$m PACS fluxes represent upper limits. The black triangle shows our LABOCA $870\,\mu$m flux determined in \cite{carina-laboca}.
The blue diamonds are the data from \cite{carina-etacar-sed}. The
dashed line represents the sum of two modified Planck spectra for dust temperatures of 210~K and 430~K from the model of \cite{carina-etacar-sed}, the solid line the sum of three Planck spectra for dust temperatures of 140~K, 200~K, and 400~K from the model of \cite{eta-car-homunculus-mass-energy}. We note that for wavelengths above $\ge 500\,\mu$m the SED gets dominated by free-free emission of the Homunculus nebula and the radiation originating
in the ionized stellar wind of $\eta$~Car.}
\label{img:eta_car_sed}
\end{figure}

Several mid- to far-infrared  observations of $\eta$~Car have been presented in the
literature, and the SED was often fitted by the sum of modified black-body
curves for different dust temperature.
In Fig.~\ref{img:eta_car_sed}, we compare our \textit{Herschel} fluxes to the
measurements and SED model described in \cite{carina-etacar-sed}. Our \textit{Herschel} SPIRE
fluxes  are well consistent (within the $1\sigma$ uncertainty range)
with this SED model prediction. In their SED model \cite{carina-etacar-sed} use the far-infrared fluxes reported by \cite{eta-car-iso-spectrum}. They were obtained with the
\textit{Kuiper Airborne Observatory} in March 1977,
i.e.~33.8 years before our \textit{Herschel} observations. This good agreement would seem to
suggest long-term stability of the thermal dust emission.

\cite{eta-car-torus-nature} presented the $2\,\mu$m to $200\,\mu$m spectrum of $\eta$~Car as
obtained in January 1996 with the \textit{Infrared Space Observatory} SWS and LWS spectrometers.
They found considerably higher far-infrared fluxes than the values of \cite{eta-car-iso-spectrum} and attributed
``these differences as a matter of calibration of the older photometry, based on
observations of Uranus as the photometric standard''.

We show in Fig.~\ref{img:eta_car_sed} the SED fit to the ISO spectrum derived by
\cite{eta-car-homunculus-mass-energy}, which is based on the sum of three modified
Planck functions with temperatures of 140~K, 200~K, and 400~K.
Compared to this SED model, our  \textit{Herschel} SPIRE fluxes are considerably lower:
The observed $250\,\mu$m flux (302~Jy) amounts to just 50\% of the level expected from
the model SED based on the ISO spectrum (624~Jy).
In the  $350\,\mu$m band, the observed flux (138~Jy) is just 57\% of the
 level expected from the model SED based on the ISO spectrum (241~Jy).
Even in the  $500\,\mu$m band, where the SED already contains a significant contribution
from free-free emission above the thermal dust emission,
the observed flux of 72~Jy is lower than the expectation from the
 model SED based on the ISO spectrum (86~Jy).
These discrepancies suggest a considerable
decrease of the far-infrared luminosity, by about a factor of two, over the last 15 years.

What could be the reason for such a drop in the far-infrared emission?
The first possibility we consider here is the
dynamical expansion of the envelope. It is known that the material in the Homunculus
moves outwards at about 600~km/s. With increasing distance from the illuminating source,
the dust is less strongly heated, thus cools down  and will produce less far-infrared emission.
Within 15 years, the outer edge of the Homunculus (where the rather cool dust that emits most far-infrared radiation is located) moves from a radial angular distance of about $9''$ to about $10''$,
and this implies that the heating of the dust at the outer edge of the
Homunculus (due to the irradiation from the central binary system)
drops by about 20\%.
This effect is too small to explain a drop of the far-infrared emission by a factor of $\sim 2$.

A more promising explanation may be
dynamical changes in the structure of the inner dust envelope around $\eta$~Car.
In the optical band, the brightness of
$\eta$~Car increased by several magnitudes during the last $\sim 30$~years
\citep[see][]{carina-eta-car-optical-lightcurve-longterm-monitoring,
carina-eta-car-submm-variability,carina-eta-car-revised-historical-lightcurve}.
This increasing optical brightness suggests that the inner envelope, that
enshrouds the central star,
is currently opening up \citep[][]{eta-car-hst-chrysalis}, and a larger fraction of the stellar
optical and UV radiation, that was previously absorbed within the nebula and
thus heated the dust, is now able to leave the system.
As a consequence, the fraction of the stellar radiation that is absorbed and heats the
dust in the envelope decreases. This finally leads to
lower levels of thermal dust emission at far-infrared wavelengths and might explain the
apparent drop of the far-infrared fluxes.

\section{Summary and Conclusions}

Our \textit{Herschel} far-infrared ($70 - 500\,\mu$m) maps of the Carina Nebula complex
revealed 642 reliable point-like sources, detected independently in at least two of the five bands.
These objects trace the youngest population of currently forming stars
in the molecular clouds. The comparison of our detection limits to models of
YSOs in different evolutionary stages shows that we can detect Class~0 protostars
(YSOs with dense envelopes)
down to stellar masses of $\sim 1\,M_\odot$, whereas for objects in later evolutionary phases
(young stars surrounded by circumstellar disks)
the limit in stellar mass is higher, $\ga 3-5\,M_\odot$.
For those 80 \textit{Herschel}-detected objects in the Carina Nebula
that can be reliably identified with an apparently single \textit{Spitzer}
counterparts, we constructed and analyzed the near-infrared to far-infrared SED
to constrain the stellar and circumstellar parameters.
About 75\% of these objects can be classified as Class~0 protostars,
based on the ratio of their sub-mm to total luminosity.
The fraction of Class~0 protostars is probably even higher among the \textit{Herschel} sources
without a clear \textit{Spitzer} counterpart.
From the number and properties of the \textit{Herschel}-detected YSOs we estimate a current star formation rate of the Carina Nebula Complex of $\sim 0.017\,M_\odot/{\rm year}$.

The SED analysis also shows that all of the 71 point-like sources with good SED fits
are low- to intermediate-mass ($1 - 10\,M_\odot$) YSOs. Since the observed distribution of
far-infrared luminosities for the \textit{Herschel} sources without clear \textit{Spitzer} counterpart is
 quite similar to those with \textit{Spitzer} counterpart, we find no indication for
the presence of highly luminous ($L \ga 10^4\,L_\odot$), i.e.~high-mass YSOs.
This implies a clear lack of high-mass YSO ($M_\ast \ga 20\,M_\odot$), although such objects should be
easily detectable in our maps, if they existed.
Considering in detail the observational detection limits, we show that
this apparent deficit of high-mass YSOs cannot be explained as
an effect of the faster evolution of circumstellar matter around more massive stars,
since the amount of circumstellar material required for a \textit{Herschel} detection
drops very strongly with increasing stellar mass (and thus luminosity) of the YSO.
The absence of high-mass YSOs is remarkable, given the presence of a
large number ($\ge 70$) of high-mass stars in the (few Myr old)
optically visible young stellar population in the Carina Nebula.

The spatial concentration of the \textit{Herschel}-detected protostars
along the edges of irradiated clouds
suggests that the currently forming generation
of stars in the CNC is predominantly triggered by the feedback from the
numerous massive stars in the several Myr old generation.

These two aspects, i.e.~the strong feedback effects leading to triggered star formation,
and the lack of massive stars in the currently forming stellar population, are probably related.
The current episode of secondary star formation occurs in clouds that are strongly
shaped and compressed by the feedback from the massive stars in the first generation.
Therefore, the physical characteristics of the current (triggered) star formation process
are quite different from the conditions that once characterized the formation of the
older (now optically visible, several Myr old) stellar population,
that includes dozens of very high-mass stars. Some fraction of the clouds present today represent
the last remaining bits of the original clouds in which the earlier stellar generation formed.
However, a large fraction of the clouds present today
have been probably swept up by the action of the massive star feedback, and
thus represent a ``second generation'' of clouds.
Their very inhomogeneous, fractal structure seems to imply that no
coherent parts of these clouds are massive and dense enough to allow the formation of massive stars
\citep[see discussion in][]{carina-herschel-clouds,carina-laboca}.

The small-scale structure of the clouds in the CNC and the star formation processes in the
individual pillars will be topics of our ongoing investigations.
The fact that the CNC represents one of the
most massive and active known Galactic star formation complexes implies that the
 detailed studies, that are possible thanks to the moderate distance of the CNC,
can serve as an important bridge to enhance our understanding
of the yet more massive, but also much more distant, extragalactic
starburst systems like 30 Doradus.

\begin{acknowledgements}
We would like to thank our referee, M.~Povich, for his constructive comments which helped to improve this paper.

The analysis of the \textit{Herschel} data was funded by the
German Federal Ministry of Economics and Technology in the framework of the
"Verbundforschung Astronomie und Astrophysik" through the DLR grant number
50 OR 1109. Additional support came from funds from the Munich Cluster
of Excellence: “Origin and Structure of the Universe”.

The \textit{Herschel} spacecraft was designed, built, tested, and launched under
a contract to ESA managed by the Herschel/Planck Project team by an industrial
consortium under the overall responsibility of the prime contractor Thales
Alenia Space (Cannes), and including Astrium (Friedrichshafen) responsible for
the payload module and for system testing at spacecraft level, Thales Alenia
Space (Turin) responsible for the service module, and Astrium (Toulouse) responsible
for the telescope, with in excess of a hundred subcontractors.

PACS has been developed by a consortium of institutes led by MPE (Germany) and including UVIE (Austria); KU Leuven, CSL, IMEC (Belgium); CEA, LAM (France); MPIA (Germany); INAF-IFSI/OAA/OAP/OAT, LENS, SISSA (Italy); IAC (Spain). This development has been supported by the funding agencies BMVIT (Austria), ESA-PRODEX (Belgium), CEA/CNES (France), DLR (Germany), ASI/INAF (Italy), and CICYT/MCYT (Spain).
SPIRE has been developed by a consortium of institutes led by Cardiff University (UK) and including Univ. Lethbridge (Canada); NAOC (China); CEA, LAM (France); IFSI, Univ. Padua (Italy); IAC (Spain); Stockholm Observatory (Sweden); Imperial College London, RAL, UCL-MSSL, UKATC, Univ. Sussex (UK); and Caltech, JPL, NHSC, Univ. Colorado (USA). This development has been supported by national funding agencies: CSA (Canada); NAOC (China); CEA, CNES, CNRS (France); ASI (Italy); MCINN (Spain); SNSB (Sweden); STFC (UK); and NASA (USA). 

This work is based in part on observations made with the \textit{Spitzer} Space
Telescope, which is operated by the Jet Propulsion Laboratory, California
Institute of Technology under a contract with NASA.

This publication makes use of data products from the Two Micron All Sky Survey, which is a joint project of the University of Massachusetts and the Infrared Processing and Analysis Center/California Institute of Technology, funded by the National Aeronautics and Space Administration and the National Science Foundation.

This publication makes use of data products from the Wide-field Infrared Survey Explorer, which is a joint project of the University of California, Los Angeles, and the Jet Propulsion Laboratory/California Institute of Technology, funded by the National Aeronautics and Space Administration. 

This research has made use of the SIMBAD database,
operated at CDS, Strasbourg, France.

\end{acknowledgements}

\bibliographystyle{aa} 
\bibliography{literatur} 

%

\Online

\onecolumn
\begin{appendix}

\section{Photometric catalogs of the \textit{Herschel} point-like sources}

 \begin{center}
\begin{tiny}
\begin{longtable}{l r r r r r r c}
\caption{Photometric catalog of the 642 \textit{Herschel} point-like sources with fluxes detected in at least two bands.}
\label{tbl:phot-fluxes-631} \\
\hline \hline \noalign{\smallskip} Source  & ${\rm F_{70}}$ & ${\rm F_{160}}$ & ${\rm F_{250}}$ & ${\rm F_{350}}$ & ${\rm F_{500}}$ & SIMBAD identification\\
  & [Jy]  & [Jy] & [Jy] & [Jy] & [Jy] & \\ \noalign{\smallskip} \hline \noalign{\smallskip}
\endfirsthead
\caption[]{(continued from previous page).} \\
\hline \hline \noalign{\smallskip} Source  & ${\rm F_{70}}$ & ${\rm F_{160}}$ & ${\rm F_{250}}$ & ${\rm F_{350}}$ & ${\rm F_{500}}$ & SIMBAD identification\\
  & [Jy]  & [Jy] & [Jy] & [Jy] & [Jy] & \\ \noalign{\smallskip} \hline \noalign{\smallskip}
\endhead
\hline
\endfoot
\hline
\endlastfoot
J103103.5--584650 & 6.39 & 8.39 & -- & -- & -- &  IRAS 10291--5831\\
J103156.1--593900 & 17.5 & 32.5 & -- & -- & -- &  IRAS 10300--5923\\
J103221.4--594235 & 6.69 & 14.6 & -- & -- & -- & \\
J103231.1--594027 & 6.46 & 7.50 & -- & -- & -- & \\
J103241.1--594000 & 9.21 & 11.9 & -- & -- & -- & \\
J103245.3--593852 & 13.3 & 40.8 & -- & -- & -- & \\
J103248.8--581849 & 73.2 & 63.1 & -- & -- & -- &  IRAS 10309--5803\\
J103251.7--585141 & -- & 2.96 & 3.78 & -- & -- & \\
\end{longtable}
\tablefoot{Last column gives the SIMBAD identification within $20\arcsec$ of the \textit{Herschel} source or the jet arising from this \textit{Herschel} source as analyzed in \cite{carina-jets-henrike}.}
\end{tiny}
\end{center}

\newpage
 \section{Photometric catalog of the point-like sources with an SED fit}
\begin{center}
\begin{tiny}
\begin{longtable}{@{}l r r r r r r r r r r r r r r@{}}
\caption{\textit{2MASS}, \textit{Spitzer}, \textit{WISE}, and \textit{Herschel} fluxes of the point-like sources with SED fits.}
\label{tbl:phot-fluxes-201} \\
\hline \hline \noalign{\smallskip} Source  & ${\rm F_{1.2}}$  &   ${\rm F_{1.7}}$  &  ${\rm F_{2.2}}$  & ${\rm F_{3.6}}$  &   ${\rm F_{4.5}}$  &   ${\rm F_{5.8}}$  &  ${\rm F_{8.0}}$ &  ${\rm F_{12}}$ &  ${\rm F_{22}}$ & ${\rm F_{70}}$ & ${\rm F_{160}}$ & ${\rm F_{250}}$ & ${\rm F_{350}}$ & ${\rm F_{500}}$ \\
  & [mJy]  & [mJy] & [mJy] & [mJy] & [mJy]  & [mJy] & [mJy] & [mJy] & [mJy] & [Jy]  & [Jy] & [Jy] & [Jy] & [Jy] \\ \noalign{\smallskip} \hline \noalign{\smallskip}
\endfirsthead
\caption[]{(continued from previous page).} \\
\hline \hline \noalign{\smallskip} Source  & ${\rm F_{1.2}}$  &   ${\rm F_{1.7}}$  &  ${\rm F_{2.2}}$  & ${\rm F_{3.6}}$  &   ${\rm F_{4.5}}$  &   ${\rm F_{5.8}}$  &  ${\rm F_{8.0}}$ &  ${\rm F_{12}}$ &  ${\rm F_{22}}$ & ${\rm F_{70}}$ & ${\rm F_{160}}$ & ${\rm F_{250}}$ & ${\rm F_{350}}$ & ${\rm F_{500}}$ \\
  & [mJy]  & [mJy] & [mJy] & [mJy] & [mJy]  & [mJy] & [mJy] & [mJy] & [mJy] & [Jy]  & [Jy] & [Jy] & [Jy] & [Jy] \\ \noalign{\smallskip} \hline \noalign{\smallskip}
\endhead
\hline
\endfoot
\hline
\endlastfoot
J103442.9--593220 & -- & -- & -- & -- & 1.46 & -- & 71.6 & -- & -- & 3.69 & 5.88 & 6.94 & 4.36 & -- \\ 
J103446.6--593139 & -- & -- & -- & -- & 0.368 & 0.766 & -- & 3.52 & 48.1 & 2.22 & 5.79 & 5.11 & 3.47 & 2.83 \\ 
J103653.5--600722 & -- & 0.744 & 3.20 & 18.4 & 46.7 & 107 & 175 & 294 & 936 & -- & 4.34 & 3.43 & 2.22 & -- \\ 
J103916.7--601022 & -- & 8.40 & 97.7 & -- & -- & 2160 & 1440 & 3430 & -- & 143 & 21.6 & 14.9 & 7.30 & 3.58 \\ 
J103921.3--590900 & -- & -- & -- & -- & 0.663 & -- & -- & -- & -- & -- & -- & 11.8 & 5.49 & 2.08 \\ 
J103942.4--595340 & -- & -- & 3.18 & 35.6 & 74.6 & 129 & 217 & 240 & 1440 & 5.46 & 13.9 & 15.7 & 10.0 & 4.64 \\ 
J103957.2--601442 & -- & -- & -- & 1.86 & 1.67 & -- & 51.0 & 167 & 241 & 5.96 & 9.38 & 6.85 & 4.29 & 2.19 \\ 
J104020.5--592557 & -- & -- & 1.55 & 3.78 & 3.28 & -- & -- & 40.7 & 89.5 & -- & -- & 16.4 & 6.26 & 7.16 \\
\end{longtable}
\tablefoot{Columns 2 -- 4 give the \textit{2MASS} fluxes, columns 5 -- 8 give the \textit{Spitzer} fluxes, and columns 9 and 10 the \textit{WISE} fluxes, respectively. The following 5 columns give the \textit{Herschel} fluxes.}
\end{tiny}
\end{center}

\newpage
\section{Model parameters of the YSO candidates}
  \begin{center}
\begin{tiny}
\begin{longtable}{l r @{}r r @{}r r @{}r r @{}r c r c}
\caption{Model parameters of the sources with good SED fits.}
\label{tbl:model-param-ranges} \\
\hline \hline \noalign{\smallskip} Source  & \multicolumn{2}{c}{Stellar mass} & \multicolumn{2}{c}{Disk mass} & \multicolumn{2}{c}{Envelope mass} & \multicolumn{2}{c}{Total luminosity} & best-fit & best & PCYC \\
    & \multicolumn{2}{c}{$[M_\odot]$}  & \multicolumn{2}{c}{$[0.01\,M_\odot]$} & \multicolumn{2}{c}{$[M_\odot]$} & \multicolumn{2}{c}{$[L_\odot]$} & model & $\chi_{\nu}^2$ & No \\ \noalign{\smallskip} \hline \noalign{\smallskip}
\endfirsthead
\caption[]{(continued from previous page).} \\
\hline \hline \noalign{\smallskip} Source  & \multicolumn{2}{c}{Stellar mass} & \multicolumn{2}{c}{Disk mass} & \multicolumn{2}{c}{Envelope mass} & \multicolumn{2}{c}{Total luminosity} & best-fit & best & PCYC \\
    & \multicolumn{2}{c}{$[M_\odot]$}  & \multicolumn{2}{c}{$[0.01\,M_\odot]$} & \multicolumn{2}{c}{$[M_\odot]$} & \multicolumn{2}{c}{$[L_\odot]$} & model & $\chi_{\nu}^2$ & No \\ \noalign{\smallskip} \hline \noalign{\smallskip}
\endhead
\hline
\endfoot
\hline
\endlastfoot
J103442.9--593220 & 6.6 & [5.7 -- 7.9] & 5.2 & [2.6 -- 27] & 19 & [4.4 -- 30] & 1300 & [780 -- 2700]  &    3009623&  3.79& \\ 
J103446.6--593139 & 2.3 & [0.18 -- 4.6] & 0.043 & [0.025 -- 25] & 61 & [0.52 -- 120] & 49 & [22 -- 110]  &    3008506&  2.68& \\ 
J103653.5--600722 & 3.3 & [0.18 -- 6.5] & 2.1 & [0.022 -- 50] & 5.5 & [0.12 -- 48] & 150 & [59 -- 990]  &    3006831&  1.16& \\ 
J103916.7--601022 & 9.7 & [7.7 -- 12] & 7.2 & [0.011 -- 61] & 53 & [12 -- 110] & 5300 & [2300 -- 11000]  &    3009705&  1.18& \\ 
J103921.3--590900 & 4.7 & [1.0 -- 9.3] & 0.44 & [0 -- 64] & 72 & [0.49 -- 250] & 330 & [63 -- 2200]  &    3008652&  0.01& \\ 
J103942.4--595340 & 6.6 & [1.7 -- 7.1] & 0.068 & [0.068 -- 39] & 130 & [6.1 -- 220] & 890 & [130 -- 890]  &    3010777&  2.52& 15\\ 
J103957.2--601442 & 1.9 & [1.3 -- 4.3] & 2.1 & [0.099 -- 12] & 77 & [12 -- 170] & 64 & [42 -- 140]  &    3008354&  4.30& \\ 
J104020.5--592557 &  & -- &   & -- &  & -- &  & -- & 3004453 &  5.67&  \\
\end{longtable}
\tablefoot{For every model parameter the best-fit value is given in the respective first column, followed by a range defined by the minimum and maximum value obtained from models constrained by a $\chi_{\nu}^2$ criterion (see Section~\ref{sec:Parameters of the protostar candidates obtained from SED fitting}). The sixth and seventh column give the identifier of the best-fit model and its $\chi_{\nu}^2$ value. Sources with no parameter values given do not match our $\chi_{\nu}^2$ criterion. The last column gives the matching source number of the Pan Carina YSO catalog \citep[PCYC;][]{carina-south-pillars-spitzer-protostars-intermediate}.}
\end{tiny}
\end{center}

\section{Additional \textit{Herschel} point-like sources}
 \begin{center}
\begin{tiny}
\begin{longtable}{l r r r r r r}
\caption{Catalog of the additional 500 \textit{Herschel} source-candidates. It includes the objects detected by {\tt CuTEx} in only one band, as well as a number of additional point-like sources that were found by visual inspection of the maps.}
\label{tbl:single-herschel-sources} \\
\hline \hline \noalign{\smallskip} Source  & ${\rm F_{70}}$ & ${\rm F_{160}}$ & ${\rm F_{250}}$ & ${\rm F_{350}}$ & ${\rm F_{500}}$ \\
  & [Jy]  & [Jy] & [Jy] & [Jy] & [Jy] \\ \noalign{\smallskip} \hline \noalign{\smallskip}
\endfirsthead
\caption[]{(continued from previous page).} \\
\hline \hline \noalign{\smallskip} Source  & ${\rm F_{70}}$ & ${\rm F_{160}}$ & ${\rm F_{250}}$ & ${\rm F_{350}}$ & ${\rm F_{500}}$ \\
  & [Jy]  & [Jy] & [Jy] & [Jy] & [Jy] \\ \noalign{\smallskip} \hline \noalign{\smallskip}
\endhead
\hline
\endfoot
\hline
\endlastfoot
J103038.9-583120 & 0.188 & 2.37 & -- & -- & -- & \\
J103105.0--584807 & -- & 11.2 & -- & -- & -- & \\
J103121.7-581920 & 1.70 & 4.04 & -- & -- & -- & \\
J103142.5-585108 & 1.83 & 5.51 & -- & -- & -- & \\
J103145.6--590911 & -- & 6.51 & -- & -- & -- & \\
J103156.6--593910 & 4.18 & -- & -- & -- & -- & \\
J103223.0-585148 & 3.66 & 11.2 & -- & -- & -- & \\
J103252.8--585204 & -- & 2.53 & -- & -- & -- & \\
\end{longtable}
\end{tiny}
\end{center}

\end{appendix}

\end{document}